\documentclass{jpp}
\usepackage{graphicx}
\usepackage[utf8]{inputenc}
\usepackage[T1]{fontenc}
\usepackage{amsmath}

\usepackage{bm}
\usepackage{amsmath}
\usepackage{xcolor}
\newcommand{\partder}[2]{\frac{\partial #1}{\partial #2}}
\newcommand{\der}[2]{\frac{d #1}{d #2}}
\usepackage[colorlinks=true]{hyperref}

\usepackage{cleveref}
\crefname{equation}{}{}
\crefname{figure}{figure}{figures}
\usepackage{subcaption}

\title{Adjoint approach to calculating shape gradients for three-dimensional magnetic confinement equilibria. Part II: Applications}

\author{Elizabeth J. Paul\aff{1}
  \corresp{\email{ejpaul@umd.edu}}, Thomas Antonsen, Jr. \aff{1}, Matt Landreman \aff{1}, W. Anthony Cooper \aff{2}}

\affiliation{\aff{1}Institute for Research in Electronics and Applied Physics, University of Maryland,
College Park, MD 20740, USA,
\aff{2}Swiss Alps Fusion Energy (SAFE), CH-1864 Vers l’Eglise, Switzerland}

\begin{document}

\maketitle

\begin{abstract}

The shape gradient is a local sensitivity function defined on the surface of an object which provides the change in a characteristic quantity, or figure of merit, associated with a perturbation to the shape of the object. The shape gradient can be used for gradient-based optimization, sensitivity analysis, and tolerance calculations. However, it is generally expensive to compute from finite-difference derivatives for shapes that are described by many parameters, as is the case for stellarator geometry. In an accompanying work \citep{Antonsen2019}, generalized self-adjointness relations are obtained for MHD equilibria. These describe the relation between perturbed equilibria due to changes in the rotational transform or toroidal current profiles, displacements of the plasma boundary, modifications of currents in the vacuum region, or the addition of bulk forces. These are applied to efficiently compute the shape gradient of functions of magnetohydrodynamic (MHD) equilibria with an adjoint approach. In this way, the shape derivative with respect to any perturbation applied to the plasma boundary or coil shapes can be computed with only one additional MHD equilibrium solution. We demonstrate that this approach is applicable for several figures of merit of interest for stellarator configuration optimization: the magnetic well, the magnetic ripple on axis, the
departure from quasisymmetry, the effective ripple in the low-collisionality $1/\nu$ regime ($\epsilon_{\text{eff}}^{3/2})$ \citep{Nemov1999}, and several finite collisionality neoclassical quantities. Numerical verification of this method is demonstrated for the magnetic well figure of merit with the VMEC code \citep{Hirshman1983} and for the magnetic ripple with modification of the ANIMEC code \citep{Cooper19923d}. Comparisons with the direct approach demonstrate that in order to obtain agreement within several percent, the adjoint approach provides a factor of $\mathcal{O}(10^3)$ in computational savings. 
\end{abstract}

\section{Introduction}

While the stellarator is a promising magnetic configuration for the realization of steady-state fusion, the geometry of a stellarator must be carefully designed. This is because collisionless charged particle trajectories are not automatically confined in a stellarator, as they are in axisymmetric configurations. Consequently, the quality of confinement depends sensitively on the shape of the confining magnetic field.  To optimize a configuration, figures of merit quantifying confinement, along with other physics criteria such as magnetohydrodynamic (MHD) stability, must be considered in numerical optimization of the MHD equilibrium. These figures of merit describing a configuration depend on the shape of the outer plasma boundary or the shape of the electromagnetic coils. It is thus desirable to obtain derivatives with respect to these shapes for optimization of equilibria or identification of sensitivity information. These so-called shape derivatives can be computed by directly perturbing the shape, recomputing the equilibrium, and computing the resulting change to a figure of merit that depends on the equilibrium solution. However, this direct finite-difference approach requires recomputing the equilibrium for each possible perturbation of the shape. For stellarators whose geometry is described by a set of $N_{\Omega}\sim 10^2$ parameters, this requires $N_{\Omega}$ solutions to the MHD equilibrium equations. Despite this computational complexity, gradient-based optimization of stellarators has proceeded with the direct approach (e.g. \cite{Reiman1999,Ku2008,Proll2015}). 

The shape gradient quantifies the change in a figure of merit associated with any perturbation to a shape. Thus, if the shape gradient can be obtained, the shape derivative with respect to \textit{any} perturbation is known (more precise definitions of the shape derivative and gradient are given in \S \ref{sec:shape_calculus}). In this work, we provide demonstration of an adjoint approach for computing the shape gradient for functions of MHD equilibria that could be considered within a stellarator configuration optimization. The adjoint approach does not require direct perturbation of a shape, but rather only the solution of one additional force balance equation which depends on the figure of merit of interest. Thus, when $N_{\Omega}$ is very large, as is generally the case for stellarator geometry, the adjoint approach is very advantageous.

In an accompanying work \citep{Antonsen2019}, two adjoint relations are derived: one involving perturbations to the plasma boundary, referred to as the fixed-boundary adjoint relation, and the other involving perturbations to currents in the vacuum region, known as the free-boundary adjoint relation. These can be considered generalizations of the self-adjointness of the force operator that arises in linearized MHD \citep{Bernstein1958}. A summary of these results is presented in \S \ref{sec:adjoint_relation}. These adjoint relations are applied to obtain expressions for the shape gradient of several figures of merit in terms of solutions to an adjoint force balance equation. 

Historically, stellarator optimization has been conducted in two stages: in the first, the plasma boundary is varied to optimize an MHD equilibrium for desired physical properties \citep{Nuhrenberg1988}. As a second step, the coils are then optimized to provide the desired plasma boundary. The fixed-boundary adjoint relation provides a means to obtain the shape gradient with respect to the plasma boundary and can be used for the traditional optimization route. 
It is also advantageous to consider coupling the coil design with the physics optimization \citep{Strickler2004,Hudson2018,Drevlak2018}, with the aim of obtaining configurations which do not require overly-complex coils. The free-boundary adjoint relation allows the computation of the shape gradient of equilibrium figures of merit with respect to coil geometry, allowing for direct optimization of coils. This approach can also be used to efficiently compute coil tolerances \citep{Landreman2018}.

Although the adjoint relations are based on the equations of linearized MHD, we perform numerical calculations in this work with non-linear MHD solutions with the addition of a small perturbation. In the accompanying paper, numerical calculations of the shape gradient with the adjoint approach were obtained for simple figures of merit that did not require modification of the Variational Moments Equilibrium Code (VMEC) \citep{Hirshman1983}. In this work, we demonstrate that the adjoint approach can be used to compute the shape gradient for other figures of merit that are relevant for the optimization of stellarator equilibria. We obtain expressions for the shape gradients of the vacuum magnetic well (\S \ref{sec:vacuum_well}), magnetic ripple (\S \ref{sec:ripple}), effective ripple in the $1/\nu$ neoclassical regime \citep{Nemov1999} where $\nu$ is the collision frequency (\S \ref{sec:epsilon_eff}), departure from quasisymmetry (\S \ref{sec:quasisymmetry}), and moments of the neoclassical distribution function (\S \ref{sec:neoclassical}) in terms of the solution to an adjoint force balance equation. We present calculations of the shape gradient with the adjoint approach for the vacuum magnetic well, which does not require modification to VMEC. The calculation for the magnetic ripple is computed with a minor modification of the Anisotropic Neumann Inverse Moments Equilibrium Code (ANIMEC) \citep{Cooper19923d}. The adjoint force balance equations needed to compute the shape gradient for the other figures of merit require the addition of a bulk force that will necessitate further modification of an equilibrium or linearized MHD code. Numerical calculations for these figures of merit will, therefore, not be presented in this work.

\section{Shape calculus fundamentals}
\label{sec:shape_calculus}

We now introduce several definitions and relations from the field of shape calculus which will prove useful for calculations in this work. Consider a function, $F(S_P)$, which depends implicitly on the plasma boundary, $S_P$, through the solution to the MHD equilibrium equations with boundary condition $\textbf{B} \cdot \textbf{n}|_{S_P} = 0$ where $\textbf{n}$ is the outward unit normal on $S_P$. We define a functional integrated over the plasma volume, $V_P$,
\begin{gather}
    f(S_P) = \int_{V_P} d^3 x \, F(S_P),
    \label{eq:f}
\end{gather}
where $S_P$ is the boundary of $V_P$. Consider a vector field describing displacements of the surface, $\delta \textbf{r}$, and a displaced surface $S_{P,\epsilon} = \{ \textbf{r}_0 + \epsilon \delta \textbf{r} : \textbf{r}_0 \in S_P\}$. The shape derivative of $F$ is defined as 
\begin{gather}
    \delta F(S_P;\delta \textbf{r}) = \lim_{\epsilon \rightarrow 0} \frac{F( S_{P,\epsilon}) - F(S_P)}{\epsilon}.
\end{gather}
The shape derivative of $f$ is defined by the same expression with $F\to f$. Under certain assumptions of smoothness of $\delta F$ with respect to $\delta \textbf{r}$, the shape derivative of the volume-integrated quantity, $f$, can be written in the following way \citep{Delfour2011_4}, 
\begin{gather}
    \delta f(S_P;\delta \textbf{r}) = \int_{V_P} d^3 x \, \delta F(S_P;\delta \textbf{r}) + \int_{S_P} d^2 x \, \delta \textbf{r} \cdot \textbf{n} F.
    \label{eq:transport_theorem}
\end{gather}
The first term accounts for the Eulerian perturbation to $F$ while the second accounts for the motion of the boundary. This is referred to as the transport theorem for domain functionals and will be used throughout to compute the shape derivatives of figures of merit of interest.

According to the Hadamard-Zol\'{e}sio structure theorem \citep{Delfour2011}, the shape derivative of a functional of $S_P$ (not restricted to the form of \eqref{eq:f}) can be written in the following form,
\begin{gather}
    \delta f(S_P;\delta \textbf{r}) = \int_{S_P} d^2 x \, \delta \textbf{r} \cdot \textbf{n} \mathcal{G},
    \label{eq:shape_gradient}
\end{gather}
assuming $\delta f$ exists for all $\delta \textbf{r}$ and is sufficiently smooth. 
In the above expression, $\mathcal{G}$ is the shape gradient. This is an instance of the Riesz representation theorem, which states that any linear functional can be expressed as an inner product with an element of the appropriate space \citep{Rudin2006}. As the shape derivative of $f$ is linear in $\delta \textbf{r}$, it can be written in the form of \eqref{eq:shape_gradient}. Intuitively, the shape derivative does not depend on tangential perturbations to the surface. The shape gradient can be computed from derivatives with respect to the set of parameters, $\Omega$, used to discretize $S_P$,
\begin{gather}
    \partder{f}{\Omega_i} = \int_{S_P} d^2 x \, \partder{\textbf{r}}{\Omega_i} \cdot \textbf{n} \mathcal{G}.
    \label{eq:shape_gradient_system}
\end{gather}
For example, $\Omega = \{ R_{mn}^c, Z_{mn}^s \}$ could be assumed, where these are the Fourier coefficients in a cosine and sine representation of the cylindrical coordinates $(R,Z)$ of $S_P$.
Upon discretization of the right-hand side on a surface, the above takes the form of a linear system that can be solved for $\mathcal{G}$ \citep{Landreman2018}. However, this approach requires performing at least one additional equilibrium calculation for each parameter with a finite-difference approach.

The shape gradient can also be computed with respect to perturbations of currents in the vacuum region. We now consider $f$ to depend on the shape of a set of filamentary coils, $C = \{ C_k \}$, through a free-boundary solution to the MHD equilibrium equations. We consider a vector field of displacements to the coils, $\delta \textbf{r}_{C}$. The shape derivative of $f$ can also be written in shape gradient form,
\begin{gather}
    \delta f(C;\delta \textbf{r}_{C}) = \sum_k \int_{C_k} dl \, \textbf{S}_k \cdot \delta \textbf{r}_{C_k},
    \label{eq:coil_shape_gradient}
\end{gather}
where $\textbf{S}_k$ is the shape gradient for coil $k$, $C_k$ is the line integral along coil $k$, and the sum is taken over coils. Again, $\textbf{S}_k$ can be computed from derivatives with respect to a set of a parameters describing coil shapes, analogous to \eqref{eq:shape_gradient_system}.

To avoid the cost of direct computation of the shape gradient, we apply an adjoint approach. The shape gradient is thus obtained without perturbing the plasma surface or coil shapes directly, but instead by solving an additional adjoint equation that depends on the figure of merit of interest. We perform the calculation with the direct approach to demonstrate that the same derivative information is computed with either approach. 

\section{Adjoint relations for MHD equilibria}
\label{sec:adjoint_relation}

Here we summarize the model for perturbed MHD equilibria and the adjoint relations from the accompanying paper. Throughout we assume the existence of magnetic coordinates such that the magnetic field can be written in the contravariant form as,
\begin{gather}
    \textbf{B} = \nabla \psi \times \nabla \theta - \iota(\psi) \nabla \psi \times \nabla \zeta,
    \label{eq:magnetic_contravariant}
\end{gather}
where $2\pi\psi$ is the toroidal flux, $\theta$ is a poloidal angle, $\zeta$ is a toroidal angle, and $\iota(\psi)$ is the rotational transform. The equilibrium magnetic field, $\textbf{B}$, is assumed to be in force balance,
\begin{gather}
    \frac{\textbf{J} \times \textbf{B}}{c} = \nabla p ,
    \label{eq:force_balance}
\end{gather}
where $\textbf{J}$ is the current density, $p(\psi)$ is the plasma pressure, and $c$ is the speed of light. The current density satisfies Ampere's law,
\begin{gather}
    \nabla \times \textbf{B} = \frac{4\pi}{c} \textbf{J}.
    \label{eq:ampere}
\end{gather}
We will consider a fixed-boundary calculation such that the equilibrium equations \eqref{eq:force_balance}-\eqref{eq:ampere} are solved with a specified value of toroidal flux $2\pi \psi_0$ on a given surface $S_P$, and free-boundary calculations such that they are solved with specified currents in the vacuum region. Two free functions of flux must also be specified, which we take to be $p(\psi)$ and the rotational transform $\iota(\psi)$ or the toroidal current contained within a flux surface,
\begin{gather}
    I_T(\psi) = \frac{c}{8\pi^2} \int_0^{2\pi} d \theta \int_0^{2\pi} d \zeta \, \sqrt{g} \textbf{B} \cdot \left( \nabla \zeta \times \nabla \psi \right),
    \label{eq:toroidal_current}
\end{gather}
where the Jacobian is $\sqrt{g} = (\nabla \psi \times \nabla \theta \cdot \nabla \zeta)^{-1}$. Fixing the toroidal current is more common in the context of stellarator optimization: for $\beta = 0$, $I_T$ can be taken to vanish while at finite $\beta$ it can be computed to be self-consistent with the neoclassical bootstrap current \citep{Spong2001,Shimizu2018}. We note that specification of an equilibrium state via \eqref{eq:magnetic_contravariant}-\eqref{eq:toroidal_current} is not always possible, as magnetic surfaces may not exist and the necessary periodicity constraints on rational surfaces may not be satisfied in a general 3-dimensional system. At this point we neglect these issues and proceed assuming that \eqref{eq:magnetic_contravariant}-\eqref{eq:toroidal_current} are sufficient to define an equilibrium state.

We now consider a linearization about this equilibrium state resulting from a perturbation to the plasma boundary, $S_P$,  the coil shapes, the scalar profiles ($p(\psi)$, $\iota(\psi)$, or $I_T(\psi)$), or the addition of a bulk force. From \eqref{eq:magnetic_contravariant}, the perturbed magnetic field can be expressed in terms of the perturbations to the magnetic coordinates coordinates, ($\delta \psi$, $\delta \theta$, $\delta \zeta$),
\begin{align}
   \delta \textbf{B} = \nabla \times [\delta \psi \left(\nabla \theta - \iota(\psi) \nabla \zeta \right) + \nabla \psi \left( \iota(\psi) \delta \zeta - \delta \theta \right)  - \delta \Phi(\psi) \nabla \zeta ],
    \label{eq:delta_B}
\end{align}
where $\delta \Phi(\psi)$ is the perturbation to the poloidal flux profile such that $\delta \Phi'(\psi) = \delta \iota(\psi)$ is the perturbation to the rotational transform profile. We can express \eqref{eq:delta_B} in terms of the displacement vector, 
\begin{align}
    \delta \textbf{B} = \nabla \times \left( \bm{\xi} \times \textbf{B} - \delta \Phi(\psi) \nabla \zeta \right),
\end{align}
with
\begin{align}
    \bm{\xi} = \frac{\textbf{B}}{B^2} \times \left[\delta \psi (\nabla \theta - \iota(\psi)\nabla \zeta) + \nabla \psi (\iota(\psi)\delta \zeta - \delta \theta ) \right]. 
    \label{eq:displacement}
\end{align}
For perturbations which fix the rotational transform profile, the familiar expression for the perturbed magnetic field from ideal MHD stability theory is recovered. 

We define a vector field which defines the displacement of a field line, $\delta \textbf{r}$, such that the perturbation to the field line label $\alpha = \theta - \iota(\psi) \zeta$ and toroidal flux satisfy,
\begin{align}
    \delta \psi + \delta \textbf{r} \cdot \nabla \psi &= 0 \\
    \delta \alpha + \delta \textbf{r} \cdot \nabla \alpha &= 0,
\end{align}
and $\delta \textbf{r} \cdot \textbf{B} = 0$. Noting that $\delta \alpha = \delta \theta - \iota(\psi) \delta \zeta - \left( \iota'(\psi) \delta \psi + \delta \Phi'(\psi) \right)\zeta $, we find that 
\begin{align}
    \delta \textbf{r} &= \bm{\xi} + \frac{\textbf{b} \times \nabla \Phi(\psi)}{B} \zeta. 
    \label{eq:delta_r}
\end{align}

The linearized force balance equation is,
\begin{gather}
    \frac{\delta \textbf{J}_{1,2} \times \textbf{B} + \textbf{J} \times \delta \textbf{B}_{1,2}}{c} - \nabla \delta p_{1,2} + \delta \textbf{F}_{1,2} = 0,
    \label{eq:perturbed_force_balance}
\end{gather}
where $\delta \textbf{J}_{1,2} = (c/4\pi) \nabla \times \delta \textbf{B}_{1,2}$ is the perturbed current density,  $\delta p_{1,2}$ is the perturbed pressure, and $\delta \textbf{F}_{1,2}$ is an additional bulk force. For perturbations that fix the pressure profile, $p(\psi)$, the change to the pressure at fixed position is 
\begin{gather}
\delta p_{1,2} = - \bm{\xi}_{1,2} \cdot \nabla p,
\end{gather}
which follows from \eqref{eq:displacement}. Quantities with subscript 1 are called the direct perturbation, and those with subscript 2 are called the adjoint perturbation. Direct perturbations correspond to a specified perturbation to the boundary, $\delta \textbf{r} \cdot \textbf{n} |_{S_P} = \bm{\xi}_1 \cdot \textbf{n} \rvert_{S_P}$, or to the coil shapes, $\delta \textbf{r}_{C}$, with no additional bulk force or perturbation to the profiles. Adjoint perturbations satisfy a modified force balance equation which depends on the figure of merit of interest. We will discuss several examples of adjoint perturbations in the following Sections.

Upon application of the self-adjointness relation for the MHD force operator \citep{Bernstein1958} augmented by the introduction of the perturbed poloidal flux, the following fixed-boundary adjoint relation is obtained,
\begin{multline}
    \int_{V_P} d^3 x \, \left(- \bm{\xi}_1 \cdot  \delta \textbf{F}_2   + \bm{\xi}_2 \cdot  \delta \textbf{F}_1 \right)
    - \frac{2\pi}{c} \int_{V_P} d \psi \, \left( \delta I_{T,2}(\psi) \delta \iota_1(\psi) - \delta I_{T,1}(\psi) \delta \iota_2(\psi) \right) \\
    - \frac{1}{4\pi} \int_{S_P} d^2 x \, \textbf{n} \cdot \left( \bm{\xi}_2 \delta \textbf{B}_1 \cdot \textbf{B} - \bm{\xi}_1 \delta \textbf{B}_2 \cdot \textbf{B} \right) = 0.
    \label{eq:fixed_boundary}
\end{multline}

For perturbed MHD equilibria, the displacement vector $\bm{\xi}$ describes the motion of the boundary ($\bm{\xi} \cdot \textbf{n} |_{S_P} = \delta \textbf{r} \cdot \textbf{n} |_{S_P})$. Thus the shape derivative with respect to the plasma boundary can be expressed with the replacement $\delta \textbf{r} \rightarrow \bm{\xi}$ (Appendix C of \cite{Antonsen2019}). Therefore we see that the boundary term in \eqref{eq:fixed_boundary} is already in the form of a shape gradient \eqref{eq:shape_gradient}. The task thus remains to express the shape derivative of a given figure of merit in terms of the first term in \eqref{eq:fixed_boundary} and convert it to shape gradient form using the fixed-boundary relation.

A similar relation is obtained for perturbation of currents in the vacuum region rather than displacements of the plasma surface, 
\begin{multline}
    \int_{V_P} d^3 x \, \left(- \bm{\xi}_1 \cdot  \delta \textbf{F}_2   + \bm{\xi}_2 \cdot  \delta \textbf{F}_1 \right) + \frac{2\pi}{c} \int_{V_P} d \psi \left( \delta \Phi_1(\psi) \der{\delta I_{T,2}(\psi)}{\psi}-\delta \Phi_2(\psi) \der{\delta I_{T,1}(\psi)}{\psi} \right) \\ + \frac{1}{c} \sum_k \left( I_{C_{k}} \int_{C_k} dl \, \left( \delta \textbf{r}_{C_{1,k}}(\textbf{x}) \cdot \textbf{t} \times \delta \textbf{B}_2 -\delta \textbf{r}_{C_{2,k}}(\textbf{x}) \cdot \textbf{t} \times \delta \textbf{B}_1 \right)\right) = 0, 
    \label{eq:free_boundary}
\end{multline}
where we have made the assumption that the currents are confined to filamentary coils, and the coil shapes are perturbed without perturbations to their currents, $I_{C_k}$. Expressions which do not make these assumptions are provided in \citep{Antonsen2019}. The unit tangent vector along the coil is $\textbf{t}$. We can note that the third term in the above expression is in the form of a coil shape gradient \eqref{eq:coil_shape_gradient}. Thus, this adjoint relation can be applied by expressing the shape derivative of a figure of merit in the form of the first two terms, and the shape gradient is computed from the solution to an adjoint equation.

These relations, \eqref{eq:fixed_boundary} and \eqref{eq:free_boundary}, will now be applied to compute the shape gradients for several figures of merit with the adjoint approach.

\section{Vacuum magnetic well}
\label{sec:vacuum_well}
The averaged radial (i.e. normal to a flux surface) curvature is an important metric for MHD stability \citep{Freidberg2014}, 
\begin{gather}
    \kappa_{\psi} \equiv \left\langle \bm{\kappa} \cdot \left(\partder{\textbf{r}}{\psi} \right)_{\alpha,l}\right\rangle_{\psi} = \left \langle \frac{1}{2B^2} \left(\partder{}{\psi} \left(8\pi p + B^2 \right) \right)_{\alpha,l} \right \rangle_{\psi},
\end{gather}
where the curvature is $\bm{\kappa} = \textbf{b} \cdot \nabla \textbf{b}$, $\textbf{b} = \textbf{B}/B$ is a unit vector in the direction of the magnetic field, $\alpha = \theta - \iota(\psi)\zeta$ is a field line label such that $\textbf{B} = \nabla \psi \times \nabla \alpha$, and $l$ measures length along a field line. Subscripts in the above expression indicate quantities held fixed while computing the derivative. The flux surface average of a quantity $A$ is \begin{gather}
    \langle A \rangle_{\psi} = \frac{\int_{-\infty}^{\infty} \frac{dl}{B}\, A}{\int_{-\infty}^{\infty} \frac{dl}{B}} =  \frac{\int_0^{2\pi} d \theta \int_0^{2\pi} d \zeta \, \sqrt{g} A}{V'(\psi)}.
\end{gather}
Here $V(\psi)$ is the volume enclosed by the surface labeled by $\psi$. The average radial curvature appears in the ideal MHD potential energy functional for interchange modes, and it provides a stabilizing effect when $p'(\psi) \kappa_{\psi} < 0$. As typically $p'(\psi)<0$, $\kappa_{\psi} >0$ is desirable for MHD stability. In a vacuum field, the expression for the averaged radial curvature reduces to 
\begin{gather}
    \kappa_{\psi} = - \frac{V''(\psi)}{V'(\psi)}.
\end{gather}
Thus, as volume increases with flux, $V''(\psi)<0$ is advantageous \citep{Helander2014}. The quantity $p'(\psi)V''(\psi)$ also appears in the Mercier criterion for ideal MHD interchange stability \citep{Mercier1974}. Known as the vacuum magnetic well, $V''(\psi)$ has been employed in the optimization of several stellarator configurations (e.g. \cite{Hirshman1999,Henneberg2019}).

We consider the following figure of merit
\begin{gather}
    f_W = \int_{V_P} d\psi \, w(\psi)V'(\psi),
    \label{eq:f_w_v}
\end{gather}
where $w(\psi)$ is a radial weight function which will be chosen so that \eqref{eq:f_w_v} approximates $V''(\psi)$. This can equivalently be written as 
\begin{gather}
    f_W = \int_{V_P} d^3 x \, w(\psi).
\end{gather}

\subsection{Fixed-boundary shape gradient}

We consider perturbations about an equilibrium with fixed toroidal current. For the direct perturbation, we have,
\begin{align}
    \delta \textbf{F}_1 &= 0 \label{eq:direct_1} \\
    \bm{\xi}_1 \cdot \textbf{n} |_{S_P} &= \delta \textbf{r} \cdot \textbf{n} |_{S_P} \label{eq:direct_2} \\
    \delta I_{T,1}(\psi) &= 0 \label{eq:direct_3},
\end{align}
for a specified boundary perturbation $\delta \textbf{r} \cdot \textbf{n}$. The shape derivative of $f_W$ is computed upon application of the transport theorem \eqref{eq:transport_theorem}, noting that $\delta \psi = - \bm{\xi}_1 \cdot \nabla \psi$,
\begin{gather}
    \delta f_W(S_P;\bm{\xi}_1) = -\int_{V_P} d^3 x \,  \bm{\xi}_1 \cdot \nabla w(\psi) + \int_{S_P} d^2 x \, \, \bm{\xi}_1 \cdot \mathbf{n} w(\psi),
    \label{eq:df_W}
\end{gather}
where we have assumed $w(\psi)$ to be differentiable. We recast the first term in \eqref{eq:df_W} as a surface integral by applying the fixed-boundary adjoint relation \eqref{eq:fixed_boundary} and prescribing the adjoint perturbation to satisfy the following,
\begin{align}
    \delta \textbf{F}_2 &= -\nabla \left(\Delta_P w(\psi) \right) \label{eq:deltaF_vacuum} \\
    \bm{\xi}_2 \cdot \textbf{n}|_{S_P} &= 0 \\
    \delta I_{T,2}(\psi) &= 0.
\end{align}
Strictly speaking, the adjoint perturbation is the linear response to the bulk force \eqref{eq:deltaF_vacuum}. Rather than solve a linearized force balance equation, we note that the adjoint bulk force takes the form of the gradient of a scalar. This is implemented by perturbing the pressure profile by $\Delta_P w(\psi)$, where $\Delta_P$ is a constant chosen judiciously. Thus a small perturbation is applied to the pressure profile, the non-linear equilibrium is computed, and the change in the fields are recorded. Accordingly $\Delta_P$ must be small enough that non-linear effects are not important, but large enough that round-off error does not dominate. 

Upon application of \eqref{eq:fixed_boundary} we obtain the following expression for the shape gradient which depends on the adjoint solution, $\delta \textbf{B}_2$,
\begin{gather}
    \mathcal{G}_{W} = \left(w(\psi) + \frac{\delta \textbf{B}_2 \cdot \textbf{B}}{4\pi \Delta_P}\right)_{S_P}.
    \label{eq:g_vacuum}
\end{gather}

\begin{figure}
    \centering
    \begin{subfigure}[b]{0.49\textwidth}
    \includegraphics[trim=3cm 3cm 3cm 6cm,clip,width=1.0\textwidth]{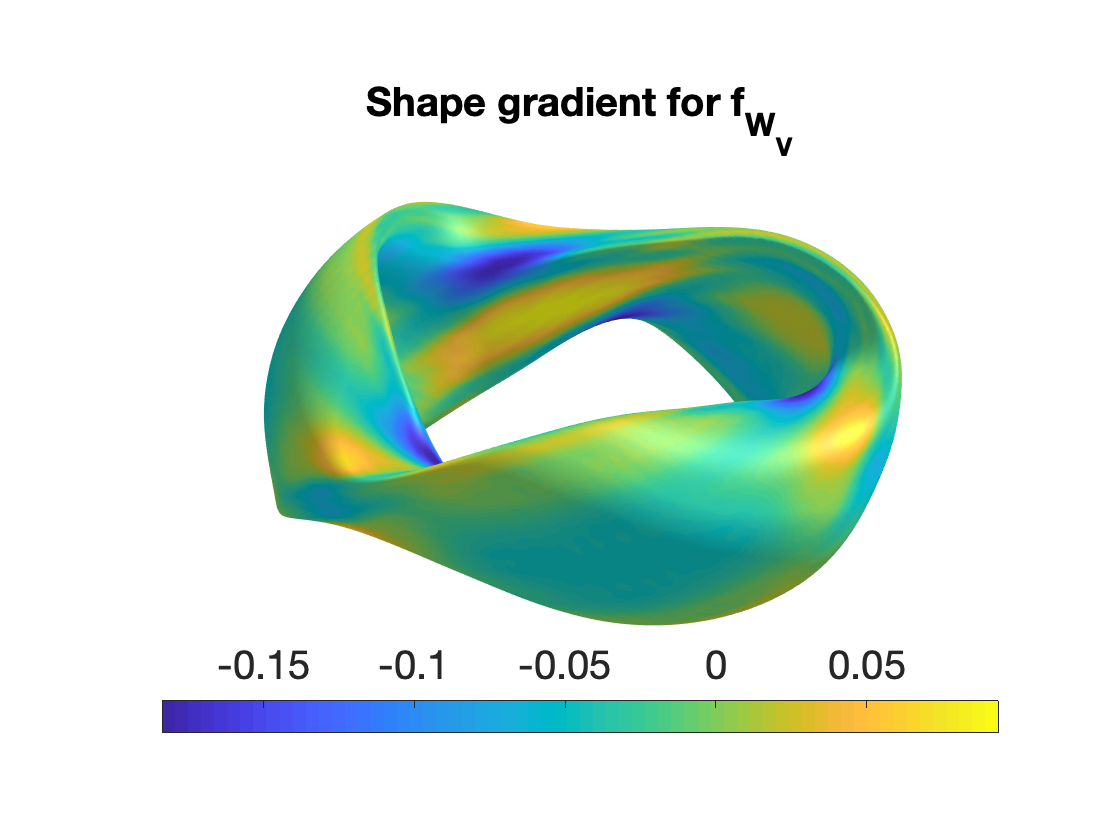}
    \caption{Adjoint}
    \label{fig:adjoint}
    \end{subfigure}
    \begin{subfigure}[b]{0.49\textwidth}
    \includegraphics[trim=3cm 3cm 3cm 6cm,clip,width=1.0\textwidth]{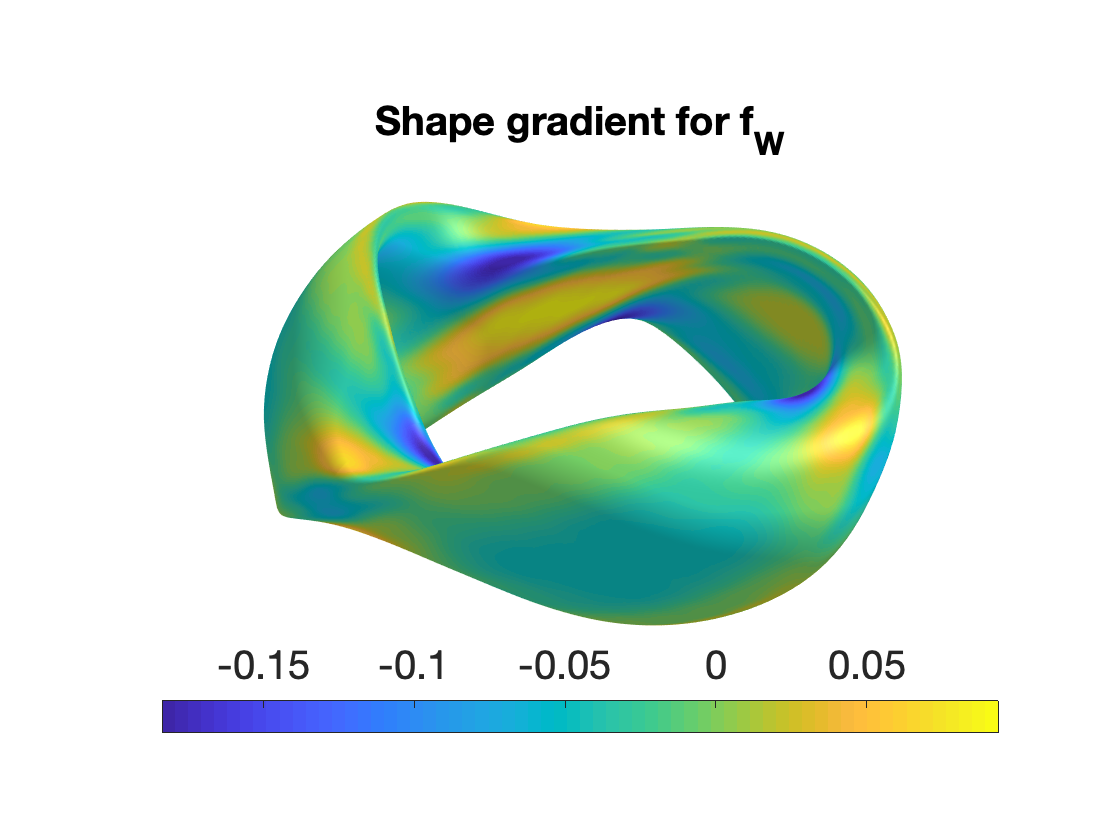}
    \caption{Direct}
    \label{fig:direct}
    \end{subfigure}
    \begin{subfigure}[b]{0.49\textwidth}
    \includegraphics[trim=1cm 6cm 1cm 6cm,clip,width=1.0\textwidth]{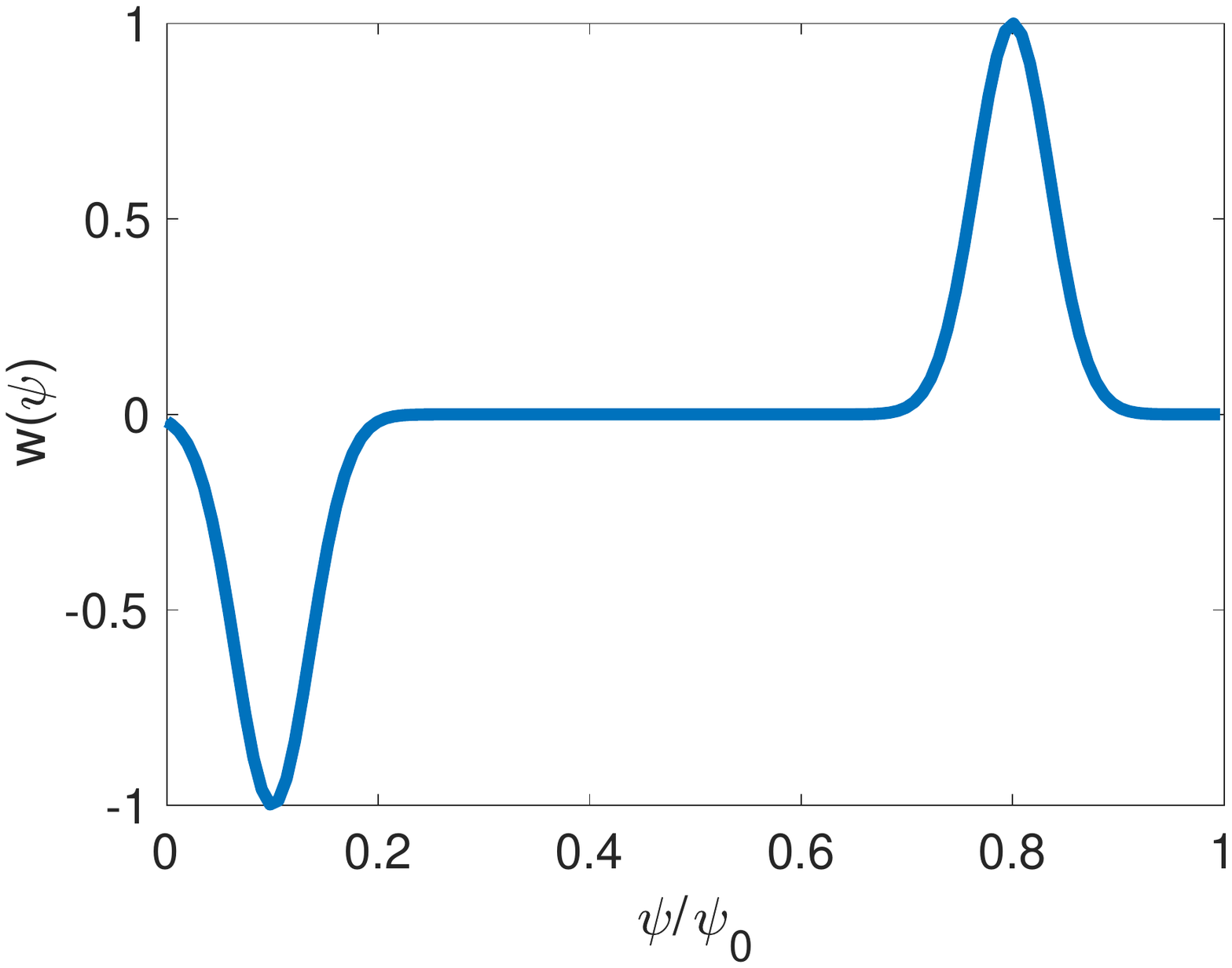}
    \caption{Weight function}
    \label{fig:weight}
    \end{subfigure}
    \caption{The shape gradient for $f_W$ \eqref{eq:f_w_v} is computed using the (a) adjoint and (b) direct approaches. (c) The weight function \eqref{eq:weight} used to compute $f_W$. }
    \label{fig:vacuum_well}
\end{figure}

In Figure \ref{fig:vacuum_well} we present the computation of $\mathcal{G}_{W}$ for the NCSX LI383 equilibrium \citep{Zarnstorff2001} using the the adjoint and direct approaches. We use a weight function 
\begin{gather}
    w(\psi) = \exp(-(\psi-\psi_{m,1})^2/\psi_w^2)- \exp(-(\psi-\psi_{m,2})^2/\psi_w^2)
    \label{eq:weight}
\end{gather}
(see Figure \ref{fig:weight})
such that $f_W$ remains smooth while it approximates $V'(\psi_{m,1})-V'(\psi_{m,2})$ where $\psi_{m,1} = 0.8 \psi_0$, $\psi_{m,2} = 0.1 \psi_0$, and $\psi_{w} = 0.05 \psi_0$. We note that $f_W$ can be interpreted as measuring the change in volume due to the interchange of two flux tubes centered at $\psi_{m,1}$ and $\psi_{m,2}$. If $f_W>0$, this indicates that moving a flux tube radially outward will cause it to expand and lower the potential energy.

All equilibrium calculations are performed with the VMEC code. For the direct approach, derivatives with respect to the Fourier discretization of the boundary ($R_{mn}^c$ and $Z_{mn}^s$) are computed for $m \le 20$ and $|n| \le 10$ using an 8-point centered difference stencil with a polynomial-fitting technique. The direct approach requires 6889 calls to VMEC while the adjoint approach requires two calls. It is clear from Figure \ref{fig:vacuum_well} that the adjoint approach yields the same gradient information as the finite-difference approach, at much lower computational cost. The small difference between Figures \ref{fig:adjoint} and \ref{fig:direct} can be quantified as follows,
\begin{gather}
S_{\text{residual}} = \frac{|S_{\text{adjoint}}-S_{\text{direct}}|}{\sqrt{\int_{S_P} d^2 x \, S_{\text{adjoint}}^2/\int_{S_P} d^2 x }}. 
\label{eq:residual}
\end{gather}
The surface-averaged value of $S_{\text{residual}}$ is $3.8\times 10^{-2}$. We note that the number of required equilibrium calculations for the direct shape gradient calculation depends on the Fourier resolution and finite-difference stencil chosen. In this work we present the number of function evaluations required in order for the adjoint and direct shape gradient calculations to agree within a few percent. As the Fourier resolution is increased, the results of the adjoint and direct methods converge to each other.

The residual difference is nonzero due to several sources of error, including discretization error in VMEC. As a result of the assumption of nested magnetic surfaces, MHD force balance \eqref{eq:force_balance} is not satisfied exactly, but a finite force residual is introduced. Error is also introduced by computing $\delta \textbf{B}_2$ with the addition of a small perturbation to a non-linear equilibrium calculation rather than from a linearized MHD solution.

\subsection{Coil shape gradient} 

\begin{figure}
    \centering
    \begin{subfigure}[b]{0.49\textwidth}
    \includegraphics[trim=8cm 4cm 4cm 3cm,clip,width=1.0\textwidth]{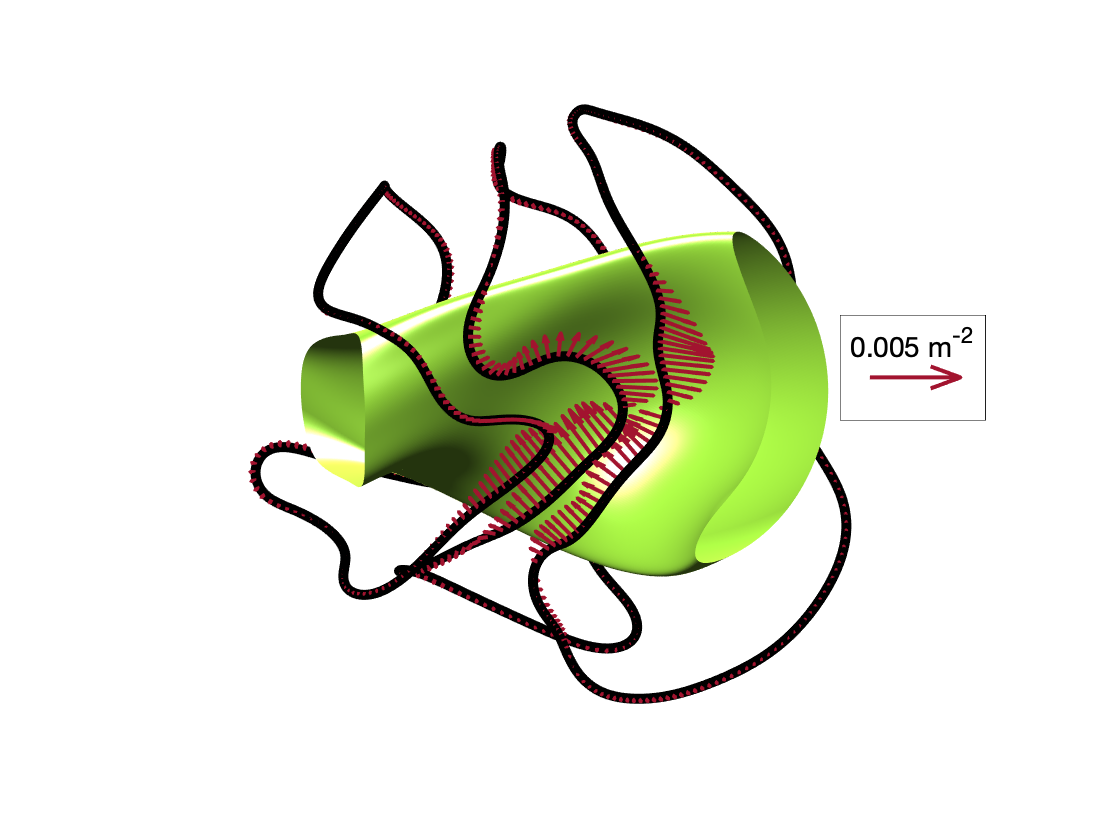}
    \caption{Adjoint}
    \end{subfigure}
    \begin{subfigure}[b]{0.49\textwidth}
    \includegraphics[trim=8cm 4cm 4cm 3cm,clip,width=1.0\textwidth]{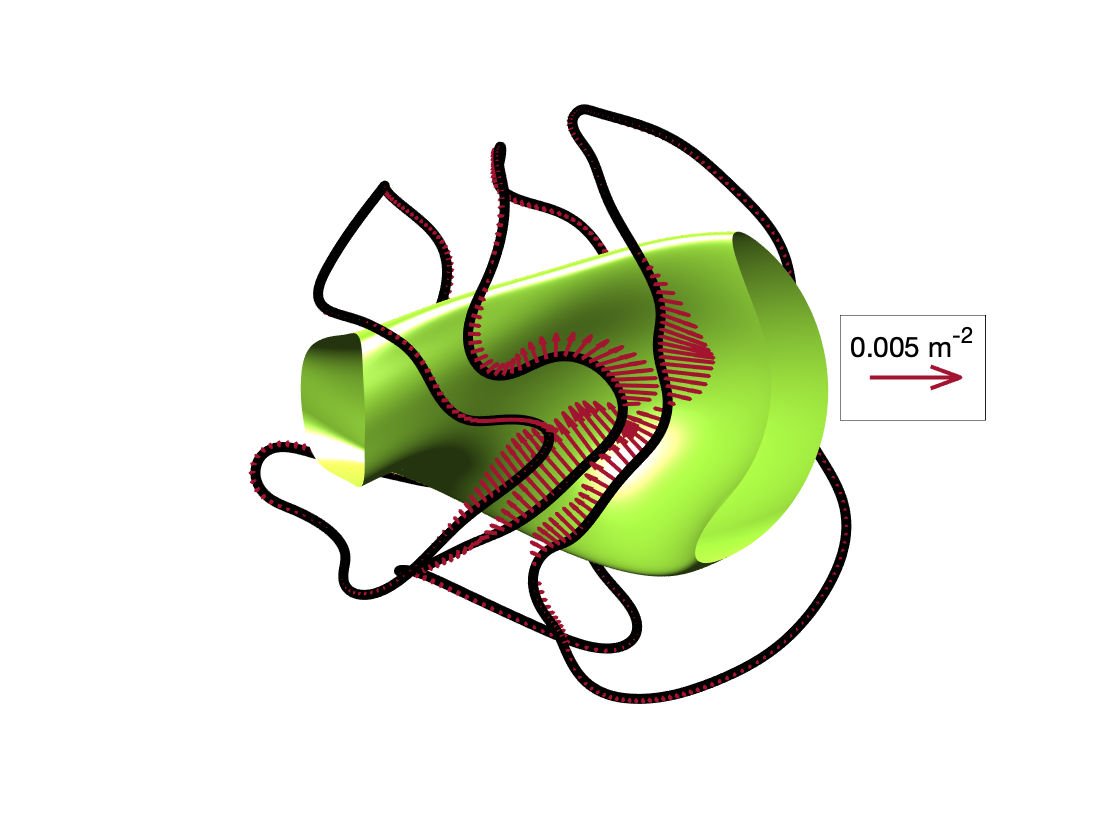}
    \caption{Direct}
    \end{subfigure}
    \caption{The coil shape gradient for $f_W$ is calculated for each of the 3 unique NCSX coil shapes. The arrows indicate the direction of $\textbf{S}_k$ \eqref{eq:well_coil_shape_gradient}, and their lengths indicate the magnitude scaled according to the legend.}
    \label{fig:coil_shape_gradient}
\end{figure}

The shape derivative of $f_W$ can also be computed with respect to a perturbation of the coil shapes. We consider perturbations about an equilibrium with fixed toroidal current, 
\begin{align}
    \delta \textbf{F}_1 &= 0 \\
    \delta I_{T,1}(\psi) &= 0,
\end{align}
with specified perturbation to the coils shapes, $\delta \textbf{r}_{C_1} \times \textbf{t}$. We prescribe the following adjoint perturbation 
\begin{align}
    \delta \textbf{F}_2 &=
    - \nabla (\Delta_P w(\psi)) \\
    \delta I_{T,2}(\psi) &= 0,
\end{align}
with $\delta\textbf{r}_{C_2} \times \textbf{t} = 0$. The same weight function \eqref{eq:weight} is applied, which decreases sufficiently fast that we can approximate $w(\psi_0) = 0$. Upon application of the free boundary adjoint relation \eqref{eq:free_boundary}, we obtain the following coil shape gradient,
\begin{gather}
    \textbf{S}_k = \frac{I_{C_k} \textbf{t} \times \delta \textbf{B}_2}{c\Delta_P} \bigg \rvert_{C_k}.
    \label{eq:well_coil_shape_gradient}
\end{gather}
The calculation of $\textbf{S}_k$ for each of the 3 unique coil shapes from the NCSX C09R00 coil set\footnote{https://princetonuniversity.github.io/STELLOPT/VMEC\%20Free\%20Boundary\%20Run} \citep{Williamson2005} is shown in Figure \ref{fig:coil_shape_gradient}. The field is computed in the vacuum region for the evaluation of $\delta \textbf{B}_2$ using the DIAGNO code \citep{Gardner1990,Lazerson2012} with a 2-point centered difference stencil. The shape gradient is also computed with a direct approach. The Cartesian components of each coil are Fourier-discretized ($\textbf{X}_m^s$,$\textbf{X}_m^c$), and derivatives are computed with respect to $m \le 40$ with a 4-point centered-difference stencil. The fractional difference between the results obtained with the two approaches is
\begin{gather}
    S_{\text{residual},k}^l = \frac{|S_{\text{adjoint},k}^l - S_{\text{direct},k}^l|}{\sqrt{\int_{C_k} dl \, \left(S_{\text{adjoint},k}^l\right)^2/\int_{C_k} dl}}.
    \label{eq:residual_coil}
\end{gather}
The line-averaged value of $S_{\text{residual},k}^l$ is $4.1\times 10^{-2}
$. The direct approach required 2917 VMEC calls while the adjoint only required three.

\section{Ripple on magnetic axis}
\label{sec:ripple}

We now consider a figure of merit which quantifies the ripple near the magnetic axis \citep{Carreras1996,Drevlak2014,Drevlak2018}.
As all physical quantities must be independent of the poloidal angle on the magnetic axis, 
this quantifies the departure from quasi-helical or quasi-axisymmetry near the magnetic axis. 

We define the magnetic ripple to be,
\begin{align}
    f_R &=  \int_{V_P} d^3 x \, \widetilde{f}_R
    \label{eq:f_R}
\end{align}
with 
\begin{subequations}
\begin{align}
   \widetilde{f_R} &= \frac{1}{2} w(\psi) \left( B - \overline{B} \right)^2 \\
   \overline{B} &= \frac{\int_{V_P} d^3 x \, w(\psi) B}{\int_{V_P} d^3 x \, w(\psi) } 
\end{align}
\end{subequations}
and a weight function given by 
\begin{align}
    w(\psi) = \exp(-\psi^2/\psi_w^2)
    \label{eq:weight_ripple}
\end{align}
with $\psi_w = 0.1 \psi_0$.

\subsection{Fixed-boundary shape gradient}

We compute perturbations about an equilibrium with fixed rotational transform
\begin{align}
    \delta \textbf{F}_1 &= 0 \\
    \bm{\xi}_1 \cdot \textbf{n} \rvert_{S_P} &= \delta \textbf{r} \cdot \textbf{n} \rvert_{S_P} \\
    \delta \iota_1(\psi) &= 0.
\end{align}
Noting that the local perturbation to the field strength is 
\begin{gather}
    \delta B = -\frac{1}{B} \left( B^2 \nabla \cdot \bm{\xi}_1 + \bm{\xi}_1 \cdot \nabla \left(B^2 + 4\pi p \right) + \delta \iota_1(\psi) \textbf{B} \cdot \left(\nabla \psi \times \nabla \zeta \right)\right),
    \label{eq:delta_mod_B}
\end{gather}
from \eqref{eq:delta_B},
the shape derivative is computed with the transport theorem \eqref{eq:transport_theorem},
\begin{gather}
        \delta f_R(S_P;\bm{\xi}_1) = \int_{S_P} d^2 x \, \bm{\xi}_1 \cdot \textbf{n} \widetilde{f_R} + \int_{V_P} d^3 x \, \left(\partder{\widetilde{f_R}}{B} \delta B + \partder{\widetilde{f_R}}{\psi} \delta \psi \right),
        \label{eq:delta_fR}
\end{gather}
where the partial derivative with respect to $B$ is performed at constant $\psi$. We prescribe the following adjoint perturbation,
\begin{align}
    \delta \textbf{F}_2 &= - \Delta_P \nabla \cdot \mathbf{P} \label{eq:ripple_F} \\
    \bm{\xi}_2 \cdot \textbf{n} |_{S_P} &= 0 \\
    \delta \iota_2(\psi) &= 0, \label{eq:ripple_iota}
\end{align}
where $\Delta_P$ is again a constant scale factor. The bulk force perturbation required for the adjoint problem is written as the divergence of an anisotropic pressure tensor, $\textbf{P} = p_{\perp} \textbf{I} + (p_{||}-p_{\perp})\textbf{b}\textbf{b}$ where $\textbf{I}$ is the identity tensor. The parallel and perpendicular pressures are related by the parallel force balance condition,
\begin{gather}
    \partder{p_{||}}{B} \bigg \rvert_{\psi} = \frac{p_{||}-p_{\perp}}{B}, 
    \label{eq:par_force_balance}
\end{gather}
which follows from the requirement that $\textbf{b} \cdot \delta \textbf{F}_2 = 0$ \eqref{eq:perturbed_force_balance}. We take the parallel pressure to be
\begin{gather}
    p_{||} = \widetilde{f_R}.
    \label{eq:p_||}
\end{gather}

Upon application of the fixed-boundary adjoint relation and the expression for the curvature in an equilibrium field, we obtain the following shape gradient,
\begin{gather}
    \mathcal{G}_R = \left( p_{\perp} +\frac{\delta \textbf{B}_2 \cdot \textbf{B}}{4\pi \Delta_P} \right)_{S_P}. 
    \label{eq:well_shape_gradient}
\end{gather}
If instead the toroidal current is held fixed in the direct perturbation as in \eqref{eq:direct_1}-\eqref{eq:direct_3}, then the required adjoint current perturbation is given by 
\begin{align}
    \delta I_{T,2}(\psi) &= \frac{c \Delta_P}{2\pi} V'(\psi)\left \langle \partder{\widetilde{f}_R}{B} \textbf{b} \cdot \nabla \zeta \times \nabla \psi \right \rangle_{\psi} \label{eq:ripple_I},
\end{align}
with the shape gradient unchanged. See Appendix \ref{app:axis_ripple} for details of the calculation. 

To compute the adjoint perturbation \eqref{eq:ripple_F}-\eqref{eq:ripple_I}, we consider the addition of an anisotropic pressure tensor to the non-linear force balance equation,
\begin{gather}
    \frac{\textbf{J}' \times \textbf{B}'}{c} = \nabla p' + \Delta_P \nabla \cdot \textbf{P}(\psi',B'),
    \label{eq:force_balance_animec}
\end{gather}
where $\textbf{P}(\psi',B') = p_{\perp}(\psi',B') \textbf{I} + \left(p_{||}(\psi',B')-p_{\perp}(\psi',B')\right)\textbf{b}' \textbf{b}'$. Here primes indicate the perturbed quantities (i.e. $B' = B + \delta B$) where unprimed quantities satisfy \eqref{eq:force_balance}. As in \S \ref{sec:vacuum_well}, the perturbation has a scale set by $\Delta_P$ which is chosen to be small enough that the response is linear. Enforcing parallel force balance from \eqref{eq:force_balance_animec} results in the following condition,
\begin{gather}
    \partder{p_{||}}{B'} \bigg \rvert_{\psi'} = \frac{p_{||}-p_{\perp}}{B'}.
    \label{eq:force_balance_||}
\end{gather}
If we furthermore assume that $\Delta_P \nabla \cdot \textbf{P}$ is small compared with the other terms in \eqref{eq:force_balance_animec}, we can consider it to be a perturbation to the base equilibrium \eqref{eq:force_balance}. In this way, we can apply the perturbed force balance equation \eqref{eq:perturbed_force_balance} with $\delta \textbf{F}_{2} = - \Delta_P \nabla \cdot \textbf{P}(\textbf{B})$, where $\textbf{P}$ is now evaluated with the equilibrium field which satisfies \eqref{eq:force_balance}. Thus the desired pressure tensor \eqref{eq:p_||} can be implemented by evaluating $p_{||}$ at the perturbed field such that \eqref{eq:force_balance_||} is satisfied. 

The pressure tensor defined by  \eqref{eq:par_force_balance}-\eqref{eq:p_||} has been implemented in the ANIMEC code \citep{Cooper19923d}, which modifies the VMEC variational principle to allow 3D equilibrium solutions with anisotropic pressures to be computed. The ANIMEC code has been used to model equilibria with energetic particle species using pressure tensors based on bi-Maxwellian \citep{Cooper2006} and slowing-down \citep{Cooper2005} distribution functions. 
The variational principle assumes that $p_{||}$ only varies on a surface through $B$ and can, therefore, be used to include the required adjoint bulk force. 

\begin{figure}
    \centering
    \begin{subfigure}[b]{0.49\textwidth}
    \includegraphics[trim=3cm 2cm 4cm 6cm,clip,width=1.0\textwidth]{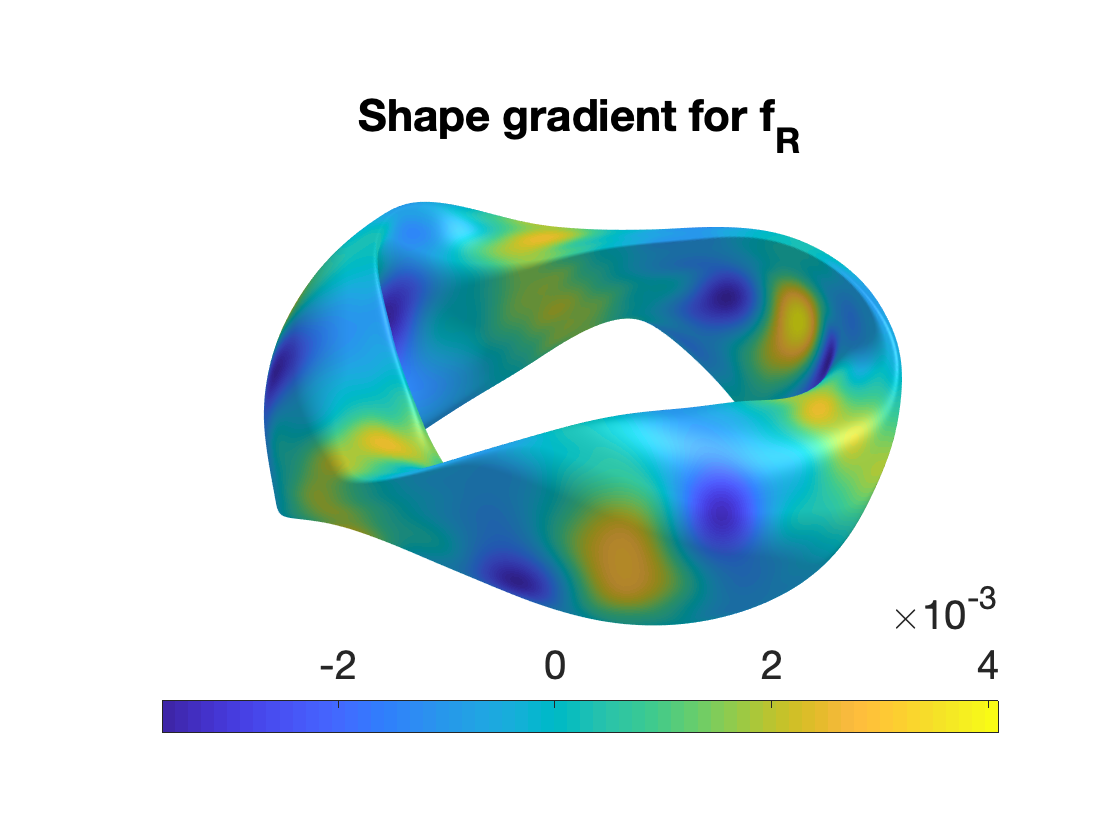}
    \caption{Adjoint}
    \end{subfigure}
    \begin{subfigure}[b]{0.49\textwidth}
    \includegraphics[trim=3cm 2cm 4cm 6cm,clip,width=1.0\textwidth]{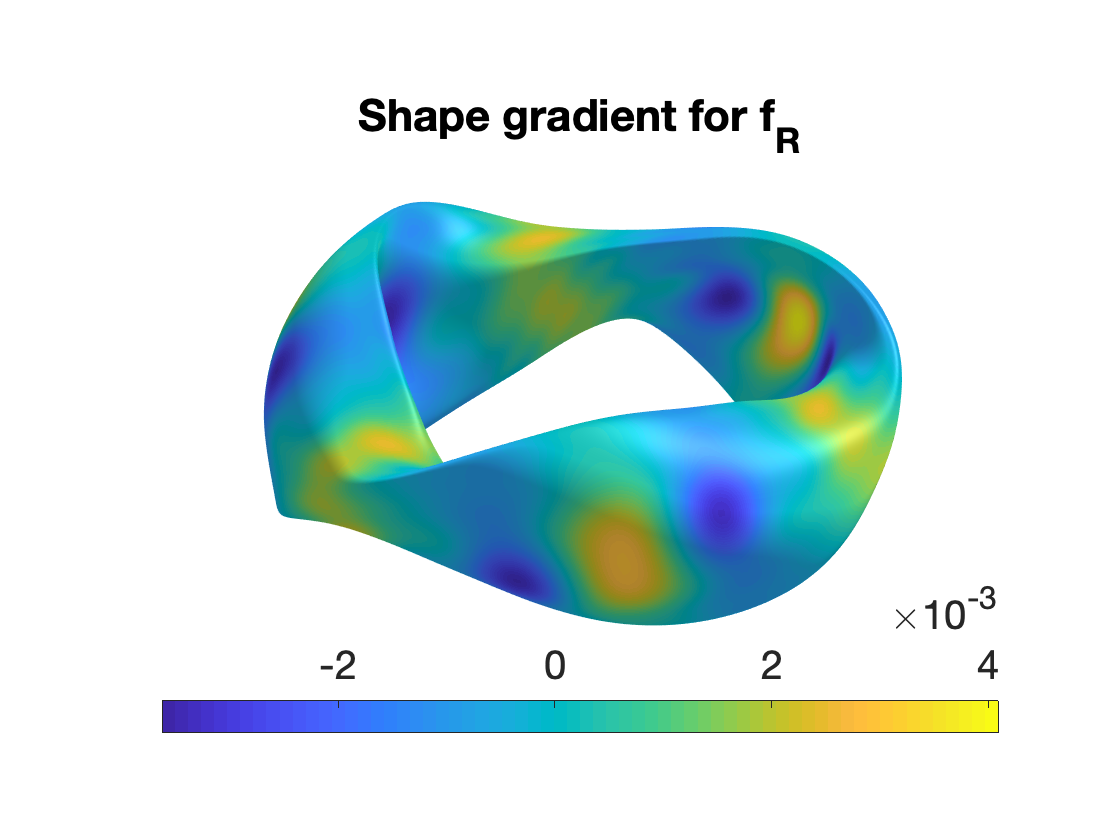}
    \caption{Direct}
    \end{subfigure}
    \begin{subfigure}[b]{0.49\textwidth}
    \includegraphics[trim=1cm 0cm 1cm 1cm,clip,width=1.0\textwidth]{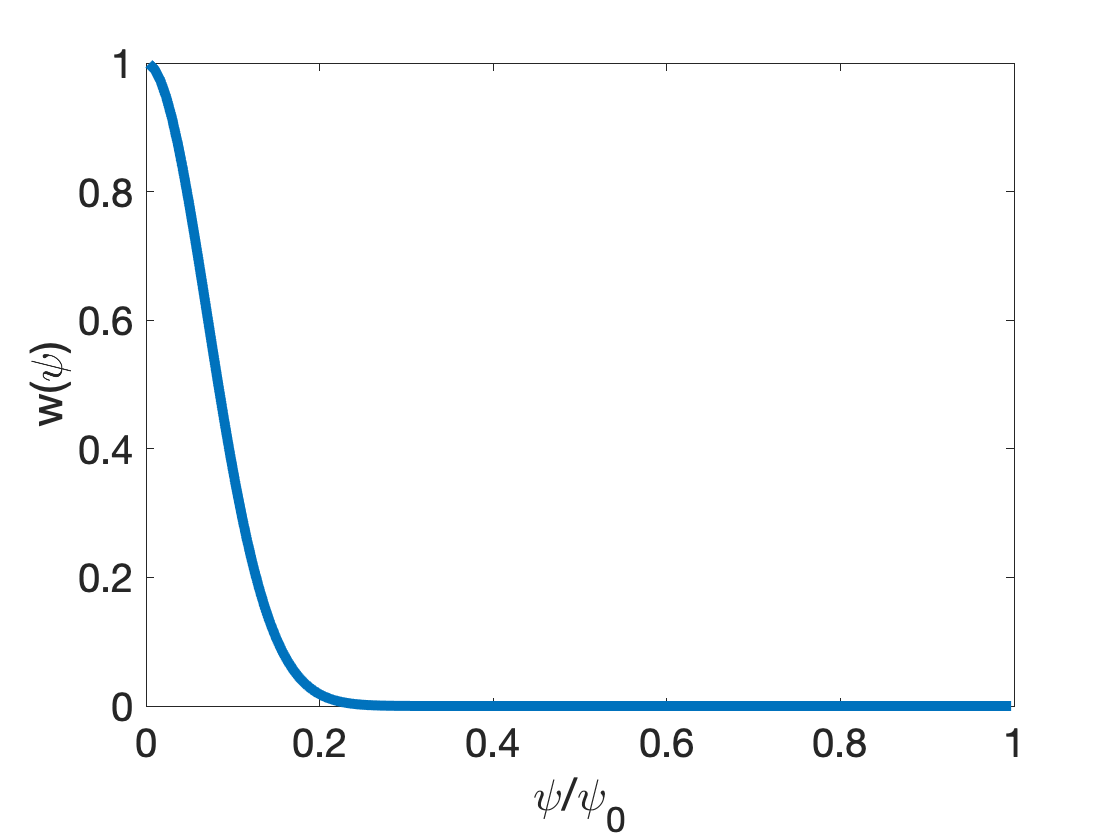}
    \caption{Weight function}
    \end{subfigure}
    \caption{The shape gradient for $f_{R}$ \eqref{eq:f_R} is computed using the (a) adjoint and (b) direct approaches with a weight function \eqref{eq:weight_ripple} shown in (c).}
    \label{fig:ripple}
\end{figure}

In Figure \ref{fig:ripple}, we present the computation of $\mathcal{G}_R$ for the NCSX LI383 equilibrium using the adjoint and direct approaches. For the direct approach, derivatives with respect to the Fourier discretization of the boundary are computed for $m \le 11$ and $|n| \le 7$ using an 8-point centered difference stencil. The direct approach required 2761 calls to VMEC while the adjoint approach required two calls. The surface-averaged value of $S_{\text{residual}}$ \eqref{eq:residual} is $3.3\times 10^{-2}$.

\section{Effective ripple in the $1/\nu$ regime}
\label{sec:epsilon_eff}

The effective ripple in the $1/\nu$ regime \citep{Nemov1999} is a figure of merit which has proven valuable for neoclassical optimization (e.g. \cite{Zarnstorff2001,Ku2008,Henneberg2019}). This quantity characterizes the geometric dependence of the neoclassical particle flux under the assumption of low-collisionality 
such that $\epsilon_{\text{eff}}$ is analogous to the helical ripple amplitude, $\epsilon_h$, that appears in the expression of the $1/\nu$ particle flux for a classical stellarator \citep{Galeev1979}. The following expression is obtained for the effective ripple,
\begin{gather}
    \epsilon_{\text{eff}}^{3/2}(\psi) = \frac{\pi}{4\sqrt{2}V'(\psi) \epsilon_{\text{ref}}^2} \int_{1/B_{\max}}^{1/B_{\min}} \frac{d \lambda}{\lambda} \,  \int_0^{2\pi} d \alpha \, \sum_i \frac{(\partder{}{\alpha}\hat{K}_i(\alpha,\lambda))^2}{\hat{I}_i(\alpha,\lambda)}.
\label{eq:eps_eff}
\end{gather}
Here $\lambda = v_{\perp}^2/(v^2B)$ is the pitch angle, $\alpha = \theta - \iota(\psi) \zeta$ is a field line label, $B_{\min}$ and $B_{\max}$ are the minimum and maximum values of the field strength on a surface labeled by $\psi$, and $\epsilon_{\text{ref}}$ is a reference aspect ratio. We have defined the bounce integrals
\begin{align}
    \hat{I}_i(\alpha,\lambda) &= \oint dl \, \frac{v_{||}}{Bv}\label{eq:I_hat} \\
    \hat{K}_i(\alpha,\lambda) &= \oint dl \, \frac{v_{||}^3}{Bv^3} \label{eq:K_hat},
\end{align}
where the notation $\oint dl = \sum_{\sigma} \sigma \int_{\zeta_-}^{\zeta_+} d \zeta/\textbf{b} \cdot \nabla \zeta$ indicates integration at constant $\lambda$ and $\alpha$ between successive bounce points where $v_{||}(\zeta_+) = v_{||}(\zeta_-) = 0$ and $\sigma = \text{sign}(v_{||})$. The sum in \eqref{eq:eps_eff} is taken over wells at constant $\lambda$ and $\alpha$ for $\zeta_{-,i} \in [0,2\pi)$.

We consider an integrated figure of merit
\begin{gather}
f_{\epsilon} = \int_{V_P} d^3 x \, w(\psi) \epsilon_{\text{eff}}^{3/2}(\psi),
\label{eq:f_epsilon}
\end{gather}
where $w(\psi)$ is a radial weight function.
We perturb about an equilibrium with fixed toroidal current \eqref{eq:direct_1}-\eqref{eq:direct_3}.
The shape derivative of $f_{\epsilon}$ is computed to be 
\begin{gather}
    \delta f_{\epsilon}(S_P;\bm{\xi}_1) = \int_{V_P} d^3 x \, \left(\textbf{P}_{\epsilon}: \nabla \bm{\xi}_1 + \delta \iota_1(\psi) \mathcal{I}_{\epsilon}\right)
    ,
    \label{eq:df_epsilon}
\end{gather}
where the double dot (:) indicates contraction between dyadic tensors $\textbf{A}$ and $\textbf{B}$ as $\textbf{A} : \textbf{B} = \sum_{i,j} A_{ij} B_{ji}$, with
\begin{multline}
\mathcal{I}_{\epsilon} = \frac{\pi w(\psi)}{2\sqrt{2} \epsilon_{\text{ref}}^2} \int_{1/B_{\max}}^{1/B} \frac{d \lambda}{\lambda} \,  \\ \times \Bigg[\frac{\left(\partder{}{\alpha} \hat{K}(\alpha,\lambda,\zeta) \right)^2}{\hat{I}^2(\alpha,\lambda,\zeta)}  \left(-\zeta \textbf{B} \times \nabla \psi \cdot \nabla  \left(\frac{|v_{||}|}{vB^2} \right) + \textbf{B} \times \nabla \psi \cdot \nabla \zeta \partder{}{B} \left( \frac{|v_{||}|}{vB} \right) \right)\\ + 2\partder{}{\alpha} \left(\frac{\partder{}{\alpha} \hat{K}(\alpha,\lambda,\zeta) }{\hat{I}(\alpha,\lambda,\zeta)} \right) \left( - \zeta \textbf{B} \times \nabla \psi \cdot \nabla \left(\frac{|v_{||}|^3}{v^3 B^2} \right) +\textbf{B} \times \nabla \psi \cdot \nabla \zeta \partder{}{B} \left(\frac{|v_{||}|^3}{v^3 B} \right) \right)\Bigg]
\end{multline}
and $\textbf{P}_{\epsilon} = p_{||} \textbf{b} \textbf{b} + p_{\perp} (\textbf{I}-\textbf{b}\textbf{b})$ with
\begin{multline}
    p_{||} = -\frac{\pi w(\psi)}{2\sqrt{2}\epsilon_{\text{ref}}^2} \int_{1/B_{\max}}^{1/B} \frac{d \lambda}{\lambda} \,  \Bigg( \frac{\left(\partder{}{\alpha} \hat{K}(\alpha,\lambda,\zeta) \right)^2}{\hat{I}^2(\alpha,\lambda,\zeta)}  \frac{|v_{||}|}{v} +2\partder{}{\alpha} \left(\frac{\partder{}{\alpha} \hat{K}(\alpha,\lambda,\zeta) }{\hat{I}(\alpha,\lambda,\zeta)} \right)\frac{|v_{||}|^3}{v^3 }\Bigg)
    \label{eq:p_perp}
\end{multline}
\begin{multline}
   p_{\perp} = - \frac{\pi w(\psi)}{2\sqrt{2}\epsilon_{\text{ref}}^2} \int_{1/B_{\max}}^{1/B} \frac{d \lambda}{\lambda} \,  \Bigg( \frac{\left(\partder{}{\alpha} \hat{K}(\alpha,\lambda,\zeta) \right)^2}{\hat{I}^2(\alpha,\lambda,\zeta)}  \left(\frac{\lambda vB}{2|v_{||}|} + \frac{|v_{||}|}{v} \right) \\ +2\partder{}{\alpha} \left(\frac{\partder{}{\alpha} \hat{K}(\alpha,\lambda,\zeta) }{\hat{I}(\alpha,\lambda,\zeta)} \right)\left( \frac{3\lambda |v_{||}|B}{2v} + \frac{|v_{||}|^3}{v^3} \right) \Bigg).
    \label{eq:p_par}
\end{multline}
Derivatives are computed assuming $\epsilon_{\text{ref}}$ is held constant. The bounce integrals are defined with respect to $\zeta$ such that $\hat{I}(\alpha,\lambda,\zeta) = \hat{I}_i$ if $\zeta \in [\zeta_{-,i},\zeta_{+,i}]$ and $\hat{I}(\alpha,\lambda,\zeta) = 0$ if $\lambda B(\alpha,\zeta) > 1$. The same convention is used for $\hat{K}(\alpha,\lambda,\zeta)$. 
We prescribe the following adjoint perturbation
\begin{align}
    \delta \textbf{F}_2 &= - \Delta_P \nabla \cdot \textbf{P}_{\epsilon} \\
    \bm{\xi}_2 \cdot \textbf{n}\rvert_{S_P} &= 0 \\
     \delta I_{T,2}(\psi) &= \frac{c  }{2 \pi} V'(\psi) \Delta_P \langle \mathcal{I}_{\epsilon}\rangle_{\psi}.
\end{align}
The adjoint bulk force must be consistent with parallel force balance from \eqref{eq:perturbed_force_balance}, which is equivalent to the condition
\begin{gather}
    \nabla_{||} p_{||} = \frac{\nabla_{||} B}{B} (p_{||}-p_{\perp}).
\end{gather}
This can be shown to be satisfied by \eqref{eq:p_perp}-\eqref{eq:p_par}, noting that the $\lambda$ integrand vanishes at $1/B$ such that there is no contribution from the parallel gradient acting on the bounds of the integral. There is also no contribution to the parallel gradient from the bounce-integrals, as $|v_{||}|$ vanishes at points of non-zero gradient of $\hat{I}(\alpha,\lambda,\zeta)$ and $\hat{K}(\alpha,\lambda,\zeta)$.

Upon application of the fixed-boundary adjoint relation \eqref{eq:fixed_boundary} and integration by parts, we obtain the following expression for the shape gradient
\begin{gather}
    \mathcal{G}_{\epsilon} = \left(p_{\perp} + \frac{\delta \textbf{B} \cdot \textbf{B}}{4\pi \Delta_P}\right)_{S_P}.
\end{gather}
See Appendix \ref{app:1_over_nu} for details of the calculation. The approach demonstrated in this Section could be extended to compute the shape gradients of other figures of merit involving bounce integrals, such as the $\Gamma_c$ metric for energetic particle confinement \citep{Nemov2005} or the variation of the parallel adiabatic invariant on a flux surface \citep{Drevlak2014}.

\section{Departure from quasisymmetry}
\label{sec:quasisymmetry}

Quasisymmetry is desirable as it ensures collisionless confinement of guiding centers. This property follows when the field strength depends on a linear combination of the Boozer angles, $B(\psi,\theta_B,\zeta_B) = B(\psi,M\theta_B-N\zeta_B)$ for fixed integers $M$ and $N$ \citep{Nuhrenberg1988,Boozer1995}. Several stellarator configurations have been optimized to be close to quasisymmetry (e.g. \cite{Reiman1999,Drevlak2013,Henneberg2019,Liu2018}) by minimizing the amplitude of symmetry-breaking Fourier harmonics of the field strength. We will consider a figure of merit that does not require a Boozer coordinate transformation; instead, we use a general set of magnetic coordinates $(\psi,\theta,\zeta)$ to define our figure of merit.

In Boozer coordinates \citep{Boozer1981,Helander2014} ($\psi,\theta_B,\zeta_B$) the covariant form for the magnetic field is 
\begin{gather}
    \textbf{B} = I(\psi) \nabla \theta_B + G(\psi) \nabla \zeta_B + K(\psi,\theta_B,\zeta_B) \nabla \psi.
    \label{eq:boozer_covariant}
\end{gather}
Here $G(\psi) = (2/c) I_P(\psi)$, where $I_P(\psi)$ is the poloidal current outside the $\psi$ surface. The poloidal current can be computed using Ampere's law and expressed as an integral over a surface labeled by $\psi$, $S_P(\psi)$,
\begin{align}
    I_P(\psi) &= \frac{c}{4\pi} \int_0^{2\pi} d \zeta \,  \textbf{B} \cdot \partder{\textbf{r}}{\zeta} \nonumber \\
    &= -\frac{c}{8\pi^2} \int_{S_P(\psi)} d^2 x \, \textbf{B} \cdot \nabla \theta \times \textbf{n}.
    \label{eq:poloidal_current}
\end{align}
The quantity $I(\psi) = (2/c) I_T(\psi)$, where $I_T(\psi)$ is the toroidal current inside the $\psi$ surface \eqref{eq:toroidal_current}. We quantify the departure from quasisymmetry in the following way,
\begin{gather}
    f_{QS} = \frac{1}{2}\int_{V_P} d^3 x \, w(\psi) \left(\textbf{B} \times \nabla \psi \cdot \nabla B -  F(\psi)\textbf{B}\cdot \nabla B\right)^2.
    \label{eq:f_QS}
\end{gather}
Here $w(\psi)$ is a radial weight function and 
\begin{gather}
    F(\psi) = \frac{(M/N)G(\psi) + I(\psi)}{(M/N)\iota(\psi)-1}.
\end{gather}
If $f_{QS} = 0$, then the field is quasisymmetric with mode numbers $M$ and $N$ \citep{Helander2014}, which can be shown using the covariant \eqref{eq:magnetic_contravariant} and contravariant \eqref{eq:boozer_covariant} representations of the magnetic field assuming $B=B(\psi,M\theta_B-N\zeta_B)$ for fixed $M$ and $N$. Note that $f_{QS}$ quantifies the symmetry in Boozer coordinates but can be evaluated in any flux coordinate system.

We consider perturbation about an equilibrium with fixed toroidal current \eqref{eq:direct_1}-\eqref{eq:direct_3}. The perturbations to the Boozer poloidal covariant component is computed using the transport theorem \eqref{eq:transport_theorem}, 
\begin{align}
     \delta G(\psi) &= -\frac{1}{4\pi^2} \int_{S_P(\psi)} d^2 x \, \left(\nabla \cdot \left( \textbf{B} \times \nabla \theta \right) \bm{\xi}_1 \cdot \textbf{n} + \delta \textbf{B} \times \nabla \theta \cdot \textbf{n} \right).
     \label{eq:delta_G_1}
\end{align}
In arriving at \eqref{eq:delta_G_1} we have used the fact that spatial derivatives commute with shape derivatives. The first term accounts for the unperturbed current density through the perturbed boundary, and the second accounts for the perturbed current density through the unperturbed boundary. The contribution from the perturbation to the poloidal angle can be shown to vanish. Upon application of \eqref{eq:delta_B} we obtain, noting that $\int_{S_P(\psi)} d^2 x \, A = V'(\psi) \langle A |\nabla \psi | \rangle_{\psi}$ for any quantity $A$,
\begin{multline}
     \delta G(\psi) = \\
     -\frac{V'(\psi)}{4\pi^2} \left \langle \bm{\xi}_1 \cdot \nabla \psi \nabla \cdot (\textbf{B} \times \nabla \theta ) - \frac{1}{\sqrt{g}} \partder{\textbf{r}}{\zeta} \cdot \nabla \times \left(\bm{\xi}_1 \times \textbf{B} \right) - \frac{\delta \iota_1(\psi)}{ \sqrt{g}^{2}} \partder{\textbf{r}}{\zeta} \cdot \partder{\textbf{r}}{\theta} \right \rangle_{\psi} \label{eq:delta_G},
\end{multline}
Applying the transport theorem \eqref{eq:transport_theorem}, the shape derivative of $f_{QS}$ takes the form, 
\begin{multline}
    \delta f_{QS}(S_P;\bm{\xi}_1) =\frac{1}{2} \int_{S_P} d^2 x \, \bm{\xi}_1 \cdot \textbf{n} \mathcal{M}^2 w(\psi) + \frac{1}{2} \int_{V_P} d^3 x \, w'(\psi) \delta \psi \mathcal{M}^2 \\
      + \int_{V_P} d^3 x \, w(\psi) \mathcal{M} \left( \delta \textbf{B} \cdot \bm{\mathcal{A}} + \bm{\mathcal{S}} \cdot \nabla \delta B + \textbf{B} \times \nabla \delta \psi \cdot \nabla B  -  \frac{\delta G(\psi) \textbf{B} \cdot \nabla B}{\iota(\psi)-(N/M)}  \right) 
      \\ + \int_{V_P} d^3 x \, w(\psi) \mathcal{M} \left( \frac{F(\psi)}{\iota(\psi)-(N/M)} \delta \iota_1(\psi) \textbf{B} \cdot \nabla B -  \delta \psi F'(\psi)\textbf{B} \cdot \nabla B \right),
      \label{eq:df_QS1}
\end{multline}
where $\mathcal{M} = \textbf{B} \times \nabla \psi \cdot \nabla B - F(\psi) \textbf{B} \cdot \nabla B$, $\bm{\mathcal{A}} = \nabla \psi \times \nabla B - F(\psi) \nabla B$, and $\bm{\mathcal{S}} = \textbf{B} \times \nabla \psi - F(\psi) \textbf{B}$. After several steps outlined in Appendix \ref{app:qs}, the shape derivative can be written in the following way,
\begin{gather}
    \delta f_{QS}(S_P;\bm{\xi}_1) = \int_{V_P} d^3 x \, \left(\bm{\xi}_1 \cdot \bm{\mathcal{F}}_{QS} + \delta \iota_1(\psi) \mathcal{I}_{QS} \right) + \int_{S_P} d^2 x \, \bm{\xi}_1 \cdot \textbf{n} \mathcal{B}_{QS} 
    \label{eq:df_QS}
\end{gather}
with
\begin{multline}
    \bm{\mathcal{F}}_{QS} =\frac{1}{2} \nabla_{\perp} \left(w(\psi) \mathcal{M}^2 \right)
    + \left((\textbf{b} \times \nabla \psi) \nabla_{||}B + F(\psi) \nabla_{\perp} B \right) w(\psi) \textbf{B} \cdot \nabla \mathcal{M} 
    \\ + \textbf{B} \times (\nabla \times (\nabla \psi \times \nabla B)) w(\psi) \mathcal{M} -B\nabla_{\perp} \left(w(\psi) \bm{\mathcal{S}} \cdot \nabla \mathcal{M}  \right)   + \bm{\kappa} Bw(\psi) \bm{\mathcal{S}} \cdot \nabla \mathcal{M}  \\ - \nabla \psi \nabla B \cdot \nabla \times \left(w(\psi) \mathcal{M} \textbf{B} \right) +\frac{1}{4\pi^2} \Bigg(- \nabla_{\perp} \left( \frac{w(\psi) V'(\psi) \langle \mathcal{M} \textbf{B} \cdot \nabla B \rangle_{\psi}}{(\iota(\psi)-(N/M))} \right) \left( \textbf{B} \cdot \nabla \psi \times \nabla \theta \right) \\
    +\frac{w(\psi) V'(\psi) \langle \mathcal{M} \textbf{B} \cdot \nabla B \rangle_{\psi}}{\iota(\psi)-(N/M)}\left(\nabla \psi\nabla \cdot \left( \textbf{B} \times \nabla \theta \right) - \textbf{B} \times \nabla \times \left(\nabla \psi \times \nabla \theta \right) \right) 
    \Bigg)
    \label{eq:F_QS}
\end{multline}
\begin{multline}
    \mathcal{B}_{QS} = -\frac{1}{2}w(\psi) \mathcal{M}^2 + B w(\psi) \bm{\mathcal{S}} \cdot \nabla \mathcal{M} - w(\psi) \mathcal{M} \nabla B \times \textbf{B} \cdot \nabla \psi \\ + \frac{w(\psi)V'(\psi) \langle \mathcal{M}\textbf{B} \cdot \nabla B \rangle_{\psi}}{4\pi^2(\iota(\psi)-(N/M))} \left(\textbf{B} \cdot \nabla \psi \times \nabla \theta  \right)
    \label{eq:B_QS}
\end{multline}
\begin{multline}
    \mathcal{I}_{QS} = - w(\psi) \mathcal{M} \nabla \psi \times \nabla \zeta \cdot \bm{\mathcal{A}} + w(\psi) \left(\bm{\mathcal{S}} \cdot \nabla \mathcal{M}\right) \textbf{b} \cdot \nabla \psi \times \nabla \zeta  \\ + \frac{w(\psi) \mathcal{M} \textbf{B} \cdot \nabla B  }{\iota(\psi)-(N/M)}\left(F(\psi) -\left \langle \frac{V'(\psi)}{4\pi^2\sqrt{g}^2} \partder{\textbf{r}}{\zeta} \cdot \partder{\textbf{r}}{\theta} \right \rangle_{\psi} \right). 
    \label{eq:I_QS}
\end{multline}
In \eqref{eq:F_QS}, $\nabla_{||} = \textbf{b} \cdot \nabla$ and $\nabla_{\perp} = \nabla- \textbf{b} \nabla_{||}$ are the parallel and perpendicular gradients. We note that $\bm{\mathcal{F}}_{QS}$ satisfies the parallel force balance condition ($\textbf{b} \cdot \bm{\mathcal{F}}_{QS}=0$) implied by \eqref{eq:perturbed_force_balance}.

We can now prescribe an adjoint perturbation which satisfies,
\begin{align}
    \delta \textbf{F}_2 &= \Delta_{QS} \bm{\mathcal{F}}_{QS} \\
    \bm{\xi} \cdot \textbf{n} |_{S_P} &= 0 \\
    \delta I_{T,2}(\psi) &= \frac{c \Delta_{QS}}{2\pi} V'(\psi) \langle \mathcal{I}_{QS} \rangle_{\psi}.
\end{align}
Upon application of the fixed-boundary adjoint relation we obtain the following shape gradient, \begin{gather}
    \mathcal{G}_{QS} = \left(\frac{\delta \textbf{B}_2 \cdot \textbf{B}}{4\pi \Delta_{QS}} + \mathcal{B}_{QS}\right)_{S_P}.
\end{gather}

\section{Neoclassical figures of merit}
\label{sec:neoclassical}

In \S \ref{sec:epsilon_eff}, we considered a figure of merit that quantifies the geometric dependence of the neoclassical particle flux in the $1/\nu$ regime. In applying this model, several assumptions are imposed, such as a small radial electric field, $E_r$, low collisionality, and a simplified pitch-angle scattering collision operator. In this Section, we consider a more general neoclassical figure of merit arising from a moment of the local drift kinetic equation, allowing for optimization at finite collisionality and $E_r$. It is assumed here that the collision time is comparable to the bounce time but shorter than the time needed to complete a magnetic drift orbit. Recently an adjoint method has been demonstrated for obtaining derivatives of neoclassical figures of merit \citep{Paul2019} with respect to local geometric quantities on a flux surface. The adjoint method described in this Section will extend these results, such that shape derivatives with respect to the plasma boundary can be computed.

Consider the following figure of merit,
\begin{gather}
    f_{NC} = \int_{V_P} d^3 x \, w(\psi) \mathcal{R}(\psi).
    \label{eq:f_NC}
\end{gather}
Here $\mathcal{R}(\psi)$ is a flux surface averaged moment of the neoclassical distribution function, $f_{1}$, which satisfies the local drift kinetic equation (DKE),
\begin{gather}
    (v_{||} \textbf{b} + \bm{v}_E)\cdot \nabla f_{1} - C(f_{1}) = - \bm{v}_{\text{m}} \cdot \nabla \psi \partder{f_{0}}{\psi},
    \label{eq:DKE}
\end{gather}
where $\bm{v}_E = \textbf{E}\times \textbf{B}/B^2$ is the $\textbf{E} \times \textbf{B}$ drift velocity, $\bm{v}_{\text{m}} \cdot \nabla \psi$ is the radial magnetic drift velocity \eqref{eq:radial_drift}, $f_{0}$ is a Maxwellian \eqref{eq:Maxwellian}, and $C$ is the linearized Fokker-Planck operator. For example, $\mathcal{R}$ can be taken to be the bootstrap current,
\begin{gather}
    J_{b} = \sum_s \frac{\langle B \int d^3 v \, f_{1s} v_{||} \rangle_{\psi}}{n_s\langle B^2 \rangle_{\psi}^{1/2}},
\end{gather}
where the sum is taken over species. We note that the geometric dependence that enters the DKE when written in Boozer coordinates only arises through the quantities $\{B, G(\psi), I(\psi), \iota(\psi) \}$. Thus for simplicity, Boozer coordinates will be assumed throughout this Section. 

The perturbation to $\mathcal{R}(\psi)$ at fixed toroidal current \eqref{eq:direct_1}-\eqref{eq:direct_3} can be written as,
\begin{gather}
    \delta \mathcal{R}(\psi) = \langle S_{\mathcal{R}} \delta B \rangle_{\psi} + \partder{\mathcal{R}(\psi)}{G(\psi)} \delta G(\psi)
+ \partder{\mathcal{R}(\psi)}{\iota (\psi)} \delta \iota_1 (\psi).
\end{gather}
Here $S_{\mathcal{R}}$ is a local sensitivity function which quantifies the change to $\mathcal{R}$ associated with a perturbation of the field strength $\delta B$ defined in the following way. Consider the perturbation to $\mathcal{R}$ resulting from a change in the field strength at fixed $G(\psi)$, $I(\psi)$, and $\iota(\psi)$. The functional derivative of $\mathcal{R}(\psi)$ with respect to $B(\textbf{r})$ can be expressed as,
\begin{gather}
    \delta \mathcal{R}(\delta B;B(\textbf{r})) = \left \langle S_{\mathcal{R}} \delta B(\textbf{r}) \right\rangle_{\psi}.
\end{gather}
This is another instance of the Riesz representation theorem: $\delta \mathcal{R}$ is a linear functional of $\delta B$, with the inner product taken to be the flux surface average. Thus $S_{\mathcal{R}}$ can be thought of as analogous to the shape gradient \eqref{eq:shape_gradient}.

The quantities $\{S_{\mathcal{R}},\partial \mathcal{R}(\psi)/\partial G(\psi),\partial \mathcal{R}(\psi)/\partial \iota(\psi) \}$ can be computed with a related adjoint method \citep{Paul2019} with the SFINCS code \citep{Landreman2014}. Here we consider SFINCS to be run on a set of surfaces such that \eqref{eq:f_NC} can be computed numerically. The derivatives computed by SFINCS will appear in the additional bulk force required for the adjoint perturbed equilibrium. 
The shape derivative of $f_{NC}$ can be computed on application of the transport theorem \eqref{eq:transport_theorem},
\begin{multline}
    \delta f_{NC}(S_P;\bm{\xi}_1) = \int_{S_P} d^2 x \, \bm{\xi}_1 \cdot \textbf{n} w(\psi) \mathcal{R}(\psi) + \int_{V_P} d^3 x \,  \delta \psi \partder{}{\psi} \left(w(\psi) \mathcal{R}(\psi)\right)\\
    + \int_{V_P} d^3 x \, 
    w(\psi) \left(\partder{\mathcal{R}(\psi)}{G(\psi)}\delta G(\psi)  + \partder{\mathcal{R}(\psi)}{\iota(\psi)}\delta \iota_1(\psi) + \left\langle S_R \delta B  \right\rangle_{\psi} \right).
    \label{eq:deltaf_NC}
\end{multline}
After several steps outlined in Appendix \ref{app:nc}, the shape derivative is written in the following form,
\begin{gather}
    \delta f_{NC} (S_P;\bm{\xi}_1) = \int_{V_P} d^3 x \, \left(\bm{\xi}_1 \cdot \bm{\mathcal{F}}_{NC} + \delta \iota_1(\psi) \mathcal{I}_{NC} \right) + \int_{S_P} d^3 x \, \bm{\xi}_1 \cdot \textbf{n} \mathcal{B}_{NC} 
    \label{eq:df_NC}
\end{gather}
with
\begin{align}
    \bm{\mathcal{F}}_{NC} &=  -\nabla( \mathcal{R}(\psi) w(\psi)) -\nabla \psi (\nabla \times \textbf{B}) \cdot \nabla \theta \partder{\mathcal{R}(\psi)}{G(\psi)} w(\psi) \frac{B^2 \sqrt{g}}{\langle B^2 \rangle_{\psi}} \nonumber \\
    &+ \frac{ w(\psi)}{\langle B^2 \rangle_{\psi}} \partder{\mathcal{R}(\psi)}{G(\psi)} \textbf{B} \times \nabla \times \left(\partder{\textbf{r}}{\zeta}B^2 \right) + G(\psi)B^2\nabla \left(\frac{w(\psi)}{\langle B^2 \rangle_{\psi}} \partder{\mathcal{R}(\psi)}{G(\psi)} \right)  \nonumber \\
    &- \bm{\kappa} w(\psi) S_{\mathcal{R}} B + B \nabla_{\perp} (w(\psi) S_{\mathcal{R}})
    \label{eq:F_NC} \\
     \mathcal{B}_{NC}
    &= w(\psi) \mathcal{R}(\psi)  -\frac{w(\psi) B^2}{\langle B^2 \rangle_{\psi}} \partder{\mathcal{R}(\psi)}{G(\psi)} G(\psi)- w(\psi) S_{\mathcal{R}} B \label{eq:B_NC} \\
    \mathcal{I}_{NC} &=   \partder{\mathcal{R}(\psi)}{G(\psi)} \frac{w(\psi) B^2}{\langle B^2 \rangle_{\psi}\sqrt{g}} \partder{\textbf{r}}{\zeta} \cdot \partder{\textbf{r}}{\theta} + w(\psi) \partder{\mathcal{R}(\psi)}{\iota(\psi)} - w(\psi) S_{\mathcal{R}} \textbf{b} \cdot \nabla \psi \times \nabla \zeta
    \label{eq:I_NC}.
\end{align}
The adjoint bulk force $\bm{\mathcal{F}}_{NC}$ is chosen to satisfy parallel force balance required by \eqref{eq:perturbed_force_balance}. 

We consider the following adjoint perturbation,
\begin{align}
    \delta \textbf{F}_2 &= \Delta_{NC} \bm{\mathcal{F}}_{NC} \\
    \bm{\xi}_2 \cdot \textbf{n} \rvert_{S_P} &= 0 \\
    \delta I_{T,2}(\psi) &= \frac{c\Delta_{NC}}{2\pi} V'(\psi) \langle \mathcal{I}_{NC} \rangle_{\psi}.
\end{align}
 Upon application of the fixed-boundary adjoint relation we obtain the shape gradient,
\begin{gather}
    \mathcal{G}_{NC} = \left(\mathcal{B}_{NC} + \frac{\delta \textbf{B}_2 \cdot \textbf{B}}{4\pi \Delta_{NC}}\right)_{S_P}.
\end{gather}

\section{Conclusions}

We have demonstrated that the self-adjointness relations (\S \ref{sec:adjoint_relation}) can be implemented to efficiently compute the shape gradient of figures of merit relevant for stellarator configuration optimization. The shape gradient is obtained by solving an adjoint perturbed force balance equation that depends on the figure of merit of interest. For the vacuum well parameter (\S \ref{sec:vacuum_well}), the additional bulk force required for the adjoint problem is simply the gradient of a function of flux, and so can be implemented by adding a perturbation to the pressure profile. For the magnetic ripple on axis (\S \ref{sec:ripple}), the required bulk force takes the form of the divergence of a pressure tensor that only varies on a surface through the field strength. As this type of pressure tensor is currently treated by the ANIMEC code, this adjoint bulk force is implemented with a minor modification to the code. Computing the shape gradient of $\epsilon_{\text{eff}}^{3/2}$ with the adjoint approach also requires the addition of the divergence of a pressure tensor. However, this pressure tensor varies on a surface through the field line label due to the bounce integrals that appear \eqref{eq:p_par}-\eqref{eq:p_perp}. Thus the variational principle used by the ANIMEC code cannot be easily extended for this application. Similarly, the shape gradients for the quasisymmetry (\S \ref{sec:quasisymmetry}) and neoclassical (\S \ref{sec:neoclassical}) figures of merit require an adjoint bulk force that is not in the form of the divergence of a pressure tensor. This provides an impetus for the development of a flexible perturbed MHD equilibrium code that could enable these calculations. While several 3D ideal MHD stability codes exist \citep{Anderson1990,Schwab1993,Strumberger2016}, only the CAS3D code has been modified in order to perform perturbed equilibrium calculations \citep{Nuhrenberg2003,Boozer2006}. We hope to take advantage of linear MHD calculations for future work through the proper modification of an existing code or development of a new code.

To date, the numerical verification of this adjoint approach for MHD equilibria has been performed with the VMEC and ANIMEC codes, which solve the non-linear force balance equations, \eqref{eq:force_balance} and \eqref{eq:force_balance_animec}. The adjoint perturbed force balance equation \eqref{eq:perturbed_force_balance} is approximated by adding a perturbative bulk force or toroidal current profile, whose characteristic magnitudes are scaled by the $\Delta$ constants (e.g. $\Delta_P$ in \eqref{eq:deltaF_vacuum}). As demonstrated in \cite{Antonsen2019}, these parameters must be small enough that non-linear effects do not become important yet large enough that round-off error does not dominate. This furthermore motivates the development of a perturbed equilibrium code that could eliminate this source of noise.

As demonstrated, this adjoint approach for functions of MHD equilibria is quite flexible and can be applied to many quantities of interest. Because of the demonstrated efficiency in comparison with the direct approach to computing shape gradients, we anticipate many further applications of this method.

\section*{Acknowledgements}

The authors would like to acknowledge productive discussions with Ricardo Nochetto on shape calculus. This work was supported by the US Department of Energy through grants DE-FG02-93ER-54197 and DE-FC02-08ER-54964. The computations presented in this paper have used resources at the National Energy Research Scientific Computing Center (NERSC). 

\bibliographystyle{jpp}

\bibliography{bibliography}

\appendix

\section{Details of axis ripple calculation}
\label{app:axis_ripple}

In this Appendix we compute the shape derivative of the finite pressure magnetic well figure of merit from \eqref{eq:delta_fR} and show that if we impose an adjoint perturbation of the form \eqref{eq:ripple_F}-\eqref{eq:ripple_iota}, the shape gradient is given by \eqref{eq:well_shape_gradient} with \eqref{eq:ripple_F}-\eqref{eq:ripple_iota}.

We use the expression for the perturbation to the field strength \eqref{eq:delta_mod_B} and $\delta \psi = - \bm{\xi}_1 \cdot \nabla \psi$ with \eqref{eq:delta_fR} to obtain
\begin{multline}
    \delta f_R(S_P;\bm{\xi}_1) = \int_{S_P} d^2 x \, \bm{\xi}_1 \cdot \textbf{n} \widetilde{f_R} - \int_{V_P} d^3 x \, \partder{\widetilde{f_R}}{\psi} \bm{\xi}_1 \cdot \nabla \psi 
    \\ - \int_{V_P} d^3 x \, \partder{\widetilde{f_R}}{B} \frac{1}{B} \left( B^2 \nabla \cdot \bm{\xi}_1 + \bm{\xi}_1 \cdot \nabla \left(B^2 + 4\pi p \right) + \delta \iota_1(\psi) \textbf{B} \cdot \left(\nabla \psi \times \nabla \zeta \right)\right). 
\end{multline}
The third term can be integrated by parts to obtain 
\begin{multline}
        \delta f_R(S_P;\bm{\xi}_1) = \int_{S_P} d^2 x \, \bm{\xi}_1 \cdot \textbf{n} \left(\widetilde{f_R}-\partder{\widetilde{f_R}}{B} B \right) + \int_{V_P} d^3 x \,\left( \partder{^2\widetilde{f_R}}{B\partial \psi}B - \partder{\widetilde{f_R}}{\psi}\right) \bm{\xi}_1 \cdot \nabla \psi 
    \\ + \int_{V_P} d^3 x \,\left(- \partder{\widetilde{f_R}}{B} B\bm{\xi}_1 \cdot  \bm{\kappa}  + B \partder{^2 \widetilde{f_R}}{B^2} \bm{\xi}_1 \cdot \nabla B + \delta \iota_1(\psi) \partder{\widetilde{f_R}}{B} \textbf{b} \cdot \left( \nabla \zeta \times \nabla \psi \right) \right)
\end{multline}
where the expression for the curvature in an equilibrium field has been applied. 

We compute one term that appears in the fixed-boundary adjoint relation \eqref{eq:fixed_boundary} using the prescribed adjoint bulk force perturbation \eqref{eq:ripple_F}
\begin{multline}
    \frac{1}{\Delta_P} \int_{V_P} d^3 x \, \bm{\xi}_1 \cdot \delta \textbf{F}_2 = \int_{V_P} d^3 x \, \left(\partder{^2 p_{||}}{B \partial \psi} B - \partder{p_{||}}{\psi} \right) \bm{\xi}_1 \cdot \nabla \psi \\
    +\int_{V_P} d^3 x \,\left(- \partder{p_{||}}{B} B \bm{\xi}_1 \cdot \bm{\kappa} + B  \partder{^2p_{||}}{B^2} \bm{\xi}_1 \cdot \nabla B \right), 
\end{multline}
where we have applied the parallel force balance condition \eqref{eq:par_force_balance}. 
Therefore, if we impose $p_{||} = \widetilde{f_R}$, we obtain the following expression for the shape derivative of $f_R$,
\begin{multline}
    \delta f_R(S_P;\bm{\xi}_1) = \int_{S_P} d^2 x \, \bm{\xi}_1 \cdot \textbf{n} \left( \widetilde{f_R} - \partder{\widetilde{f_R}}{B} B \right) + \frac{1}{\Delta_P} \int_{V_P} d^3 x \, \bm{\xi}_1 \cdot \delta \textbf{F}_2 \\
    + \int_{V_P} d^3 x  \, \delta \iota_1(\psi) \partder{\widetilde{f_R}}{B} \textbf{b} \cdot \left( \nabla \zeta \times \nabla \psi \right).
\end{multline}
Upon application of the fixed-boundary adjoint relation we obtain \eqref{eq:well_shape_gradient} with \eqref{eq:ripple_F}-\eqref{eq:ripple_iota}.

\section{Details of effective ripple in the $1/\nu$ regime calculation}
\label{app:1_over_nu}

Neoclassical transport in the $1/\nu$ collisionality regime is discussed in many references including \cite{Frieman1970}, \cite{Connor1974}, and \cite{Ho1987}. In this Appendix we sketch the computation of  $\epsilon_{\text{eff}}^{3/2}$ originally introduced in \cite{Nemov1999} and compute linear perturbations of $f_{\epsilon}$ \eqref{eq:f_epsilon}, showing them to take the form of \eqref{eq:df_epsilon}.

In the $1/\nu$ regime, the distribution function is ordered in the parameter $\nu_* = \nu/(v_t/L) \ll 1$, where $\nu$ is the collision frequency, the thermal speed is $v_t = \sqrt{2T/m}$ for mass $m$ and temperature $T$, and $L$ is a macroscopic scale length,
\begin{gather}
    f_{1} = f_{1}^{-1} + f_{1}^0 + \mathcal{O}(\nu_*).
\end{gather}
In velocity space we use a pitch angle coordinate
$\lambda = v_{\perp}^2/(v^2B)$, energy coordinate $\epsilon = v^2/2$, and $\sigma = \text{sign}(v_{||})$, where $v_{\perp} = \sqrt{v^2-v_{||}^2}$ is the perpendicular velocity and $v_{||} = \bm{v} \cdot \textbf{b}$ is the parallel velocity. We use the field line label, $\alpha$, and length along a field line, $l$, to describe location on a constant $\psi$ surface. In the $1/\nu$ regime the $\textbf{E}\times \textbf{B}$ precession frequency is assumed to be small relative to the collision frequency, and the drift kinetic equation is,
\begin{gather}
    v_{||} \partder{f_{1}}{l} = C(f_{1}) - \bm{v}_{\text{m}} \cdot \nabla \psi \partder{f_{0}}{\psi},
\end{gather}
where the Maxwellian with density $n$ is 
\begin{gather}
f_{0} = n \pi^{-3/2} v_{t}^{-3} e^{-v^2/v_{t}^2}
\label{eq:Maxwellian}
\end{gather}
and the radial magnetic drift is
\begin{gather}
    \bm{v}_{\text{m}} \cdot \nabla \psi = (v^2 + v_{||}^2) \frac{mc}{2q B^3} \nabla \psi \times \textbf{B} \cdot \nabla B,
    \label{eq:radial_drift}
\end{gather}
for charge $q$. The drift kinetic equation to $\mathcal{O}(\nu_*^{-1})$ is 
\begin{gather}
    v_{||} \partder{f_{1}^{-1}}{l} = 0.
\end{gather}
In the trapped portion of phase space, this implies that $f_{1}^{-1}=f_{1}^{-1}(\psi,\alpha,\epsilon,\lambda)$, and in the passing portion of phase space, this implies that $f_{1}^{-1}=f_{1}^{-1}(\psi,\epsilon,\lambda,\sigma)$. The drift kinetic equation to $\mathcal{O}(\nu_*^0)$ is
\begin{gather}
    v_{||} \partder{f_{1}^0}{l} = C(f_{1}^{-1}) - \bm{v}_{\text{m}} \cdot \nabla \psi \partder{f_{0}}{\psi}.
    \label{eq:kinetic_equation}
\end{gather}
In the passing region, this implies that $f_{1}^{-1}$ is a Maxwellian, so it can be taken to vanish. We employ a pitch-angle scattering operator,
\begin{gather}
    C = \frac{2\nu(\epsilon) v_{||}}{B\epsilon} \partder{}{\lambda} \left(\lambda v_{||}\partder{}{\lambda} \right).
\end{gather}
The parallel streaming term in \eqref{eq:kinetic_equation} is annihilated by the bounce averaging operation,  
\begin{gather}
    0 = \langle C(f_1^{-1})\rangle_{b} - \langle \bm{v}_{\text{m}} \cdot \nabla \psi\rangle_{b} \partder{f_0}{\psi}.
    \label{eq:bounce_averaged}
\end{gather}
where the bounce average of a quantity $A$ is $\langle A \rangle_b = \tau^{-1} \oint dl \, A/v_{||}$ and the bounce time is $\tau = \oint dl \, v_{||}^{-1}.$ The bounce-averaged equation \eqref{eq:bounce_averaged} can be expressed in terms of the parallel adiabatic invariant $J = \oint dl \, v_{||}$ using the relation
\begin{gather}
    \langle \bm{v}_{\text{m}} \cdot \nabla \psi \rangle_{b} = \frac{mc}{q\tau} \partder{J}{\alpha}.
\end{gather}
Integrating \eqref{eq:bounce_averaged} with respect to $\lambda$ we obtain,
\begin{gather}
    \partder{f_{1}^{-1}}{\lambda} =\frac{mc\epsilon}{2q\lambda\nu(\epsilon)} \partder{f_{0}}{\psi} \left(\oint dl \, \frac{v_{||}}{B} \right)^{-1} \int_{1/B_{\max}}^{\lambda} d \lambda' \, \partder{J}{\alpha}.
\end{gather}
Here $B_{\max}$ is the maximum value of the field strength on the surface labeled by $\psi$.
We have used the boundary condition $\left( \oint dl \, v_{||}/B \right) \partial f_1^{-1}/\partial \lambda \rvert_{\lambda = 1/B_{\max}} =0$, as there is no flux in pitch-angle from the passing region. The integration with respect to $\lambda$ is performed to obtain, 
\begin{gather}
    \partder{f_{1}^{-1}}{\lambda} = -\frac{mc}{6q\lambda \nu(\epsilon)} \partder{f_0}{\psi} \left(\oint dl \, \frac{v_{||}}{B} \right)^{-1} \partder{}{\alpha}\left(\oint dl \, \frac{v_{||}^3}{B} \right) .
    \label{eq:dfdlambda}
\end{gather}
The particle flux from $f_1^{-1}$ is obtained by multiplying \eqref{eq:kinetic_equation} by $f_{1}^{-1} (\partial f_{0}/\partial \psi)^{-1}$, integrating over velocity space, and flux surface averaging,
\begin{align}
    \left \langle \bm{\Gamma} \cdot \nabla \psi \right \rangle_{\psi} &\equiv \left \langle \int d^3 v \, f_{1}^{-1} \bm{v}_{\text{m}} \cdot \nabla \psi \right \rangle_{\psi} 
    = \left \langle \int d^3 v \, f_{1}^{-1} C(f_{1}^{-1}) \left(\partder{f_{0}}{\psi} \right)^{-1} \right \rangle_{\psi}.
\end{align}
The velocity space integration is performed using the Jacobian $d^3 v = 2\pi \sum_{\sigma} B \epsilon/|v_{||}| d \lambda d \epsilon$. Upon integration by parts in $\lambda$ and applying \eqref{eq:dfdlambda}, the following expression is obtained,
\begin{multline}
    \left \langle \bm{\Gamma} \cdot \nabla \psi \right \rangle_{\psi} = \\ - \frac{4\sqrt{2}\pi}{V'(\psi)} \left(\frac{mc}{3q} \right)^2 \int_0^{\infty} d \epsilon \, \left(\partder{f_{0}}{\psi} \right) \frac{\epsilon^{5/2}}{\nu(\epsilon)} \int_{1/B_{\max}}^{1/B_{\min}} \frac{d \lambda}{\lambda} \,  \int_0^{2\pi} d \alpha \, \sum_i \frac{(\partder{}{\alpha}\hat{K}_i(\alpha,\lambda))^2}{\hat{I}_i(\alpha,\lambda)},
    \label{eq:particle_flux}
\end{multline}
where the bounce integrals are defined by \eqref{eq:I_hat} and \eqref{eq:K_hat}. The sum in \eqref{eq:particle_flux} is taken over trapping regions for particles with pitch angle $\lambda$ on a field line labeled by $\alpha$. The sum is carried out for left bounce points $\zeta_{-,i} \in[0,2\pi)$.

The parameter $\epsilon_{\text{eff}}^{3/2}$ quantifies the geometric dependence of the $1/\nu$ particle flux.
It is defined in terms of the radial particle flux in the following way \citep{Nemov1999},
\begin{align}
 \langle \bm{\Gamma} \cdot \nabla \psi \rangle_{\psi} = -32\langle |\nabla \psi | \rangle_{\psi}^2 \left(\frac{mc}{3q}\right)^2 \frac{1}{B_0^2R^2}\epsilon_{\text{eff}}^{3/2} \int_0^{\infty} d \epsilon \, \left(\partder{f_{0}}{\psi} \right) \frac{\epsilon^{5/2}}{\nu(\epsilon)} . 
 \label{eq:classical_flux}
\end{align}
We take our normalizing length and field values to be such that $B_0 R = \epsilon_{\text{ref}}^{-1} \langle |\nabla \psi | \rangle_{\psi}$, where $\epsilon_{\text{ref}}$ is a reference aspect ratio. Comparing \eqref{eq:particle_flux} with \eqref{eq:classical_flux} we obtain the expression for $\epsilon_{\text{eff}}^{3/2}$ \eqref{eq:eps_eff}. The corresponding expression (29) in \cite{Nemov1999} is obtained by noting that $\hat{H}^{\text{Nemov}} = -(\partial \hat{K}/\partial \alpha) \lambda^{1/2} B_0^{3/2}$ and $\hat{I} = 2\hat{I}^{\text{Nemov}}$, where $\hat{H}^{\text{Nemov}}$ and $\hat{I}^{\text{Nemov}}$ are given in (30)-(31) of \cite{Nemov1999}.

The shape derivative of $f_{\epsilon}$ \eqref{eq:f_epsilon} is computed to be 
\begin{gather}
    \delta f_{\epsilon} (S_P;\bm{\xi}_1) = \int_{V_P} d \psi \,  w(\psi) \delta (V'(\psi)\epsilon_{\text{eff}}^{3/2}(\psi)).
\label{eq:df_epsilon2}
\end{gather}
The perturbation to the bounce integrals is computed using the following identity for the perturbation of a line integral $Q_L = \int_{l_0}^{l_L} dl \, Q$ due to displacement of the integration curve by vector field $\delta \textbf{r}$ \citep{Antonsen1982,Landreman2018},
\begin{gather}
    \delta Q_L = \int_{l_0}^{l_L} dl \,  \left( \delta \textbf{r} \cdot \left(-\bm{\kappa}Q + \left(\textbf{I}-\textbf{t}\textbf{t} \right)\cdot \nabla Q\right)  + \delta Q \right) + Q(l_L) \delta l_L - Q(l_0) \delta l_0,
\end{gather}
where $\delta Q$ is the perturbation to the integrand at fixed position, $\textbf{t} = \partial \textbf{r}/\partial l$ is the unit tangent vector, $\bm{\kappa} = \partial^2 \textbf{r}/\partial l^2$ is the curvature, and $\delta l_L$ and $\delta l_0$ are perturbations to the bounds of the integral.

We compute the perturbation to the bounce integrals to be 
\begin{align}
    \delta \hat{I}_i &=\oint dl \, \left(-\frac{v_{||}}{vB} \bm{\kappa} \cdot \delta \textbf{r} - \left(\frac{\lambda v}{2B v_{||}}+\frac{v_{||}}{B^2v} \right) \left(\delta \textbf{r} \cdot \nabla B + \delta B \right) \right) \\
    \delta \hat{K}_i &= \oint dl \, \left(-\frac{v_{||}^3}{v^3B} \bm{\kappa} \cdot \delta \textbf{r} - \left(\frac{3\lambda v_{||}}{2Bv} + \frac{v_{||}^3}{B^2v^3} \right)\left(\delta \textbf{r} \cdot \nabla B + \delta B \right) \right) 
\end{align}
where $\delta B$ is the perturbation to the field strength \eqref{eq:delta_mod_B} and $\delta \textbf{r}$ is given by \eqref{eq:delta_r}. We note that $\delta \textbf{r} \cdot \textbf{b} = 0$ such that the perpendicular projection, $(\textbf{I}-\textbf{t}\textbf{t})$, is not needed. There is no contribution due to the perturbation of the bounce points, as the integrand vanishes at these points. The expressions \eqref{eq:df_epsilon}-\eqref{eq:p_par} can now be obtained by writing \eqref{eq:df_epsilon2} in terms of the perturbations of the bounce integrals, using $\bm{\xi}_1 \cdot \nabla B + \delta B = - B\left(\textbf{I} - \textbf{b} \textbf{b} \right):\nabla \bm{\xi}_1 - \delta \iota_1 \textbf{b} \cdot (\nabla \psi \times \nabla \zeta)$ and 
$\bm{\kappa} \cdot \bm{\xi}_1 = - \textbf{b} \textbf{b}: \nabla \bm{\xi}_1$.

\section{Details of departure from quasisymmetry calculation}
\label{app:qs}

In this Appendix we compute the shape derivative of $f_{QS}$ \eqref{eq:f_QS} to obtain \eqref{eq:df_QS}-\eqref{eq:I_QS} by expressing each term in \eqref{eq:df_QS1} in the desired form. The second term in \eqref{eq:df_QS1} is expressed using $\delta \psi = - \bm{\xi}_1 \cdot \nabla \psi$,
\begin{gather}
    \frac{1}{2}\int_{V_P} d^3 x \, w'(\psi) \delta \psi \mathcal{M}^2 = -\frac{1}{2} \int_{V_P} d^3 x \,  \mathcal{M}^2 \bm{\xi}_1 \cdot \nabla w(\psi) .
\end{gather}
The third term in \eqref{eq:df_QS1} is computed upon application of \eqref{eq:delta_B}, the divergence theorem, and noting that $\mathcal{M} = \textbf{B} \cdot \bm{\mathcal{A}}$, 
\begin{multline}
    \int_{V_P} d^3 x \, w(\psi) \mathcal{M} \delta \textbf{B} \cdot \bm{\mathcal{A}} =  - \int_{S_P} d^2 x \,  \bm{\xi}_1 \cdot \textbf{n}  w(\psi) \mathcal{M}^2 - \int_{V_P} d^3 x \, w(\psi) \delta \iota_1(\psi) \mathcal{M} \nabla \psi \times \nabla \zeta \cdot \bm{\mathcal{A}} \\
    +\int_{V_P} d^3 x \, \bm{\xi}_1 \cdot \left(w(\psi) \mathcal{M} \left( \textbf{B} \times \left( \nabla \times \bm{\mathcal{A}} \right) \right) - \bm{\mathcal{A}} w(\psi) \textbf{B} \cdot \nabla \mathcal{M}  + \mathcal{M}\nabla \left( w(\psi)\mathcal{M} \right) \right).
    \label{eq:qs_2}
\end{multline}
The quantity $\bm{\mathcal{A}}$ can be projected into the perpendicular direction as $\bm{\xi}_1 \cdot \textbf{b} = 0$, noting that
\begin{gather}
    \textbf{b} \times \left( \bm{\mathcal{A}} \times \textbf{b} \right) = -(\textbf{b} \times \nabla \psi) \nabla_{||} B - F(\psi) \nabla_{\perp} B.
\end{gather}
Similarly, any terms in \eqref{eq:qs_2} involving $\bm{\xi}_1 \cdot \nabla$ can be expressed as $\bm{\xi}_1 \cdot \nabla_{\perp}$. The corresponding terms in \eqref{eq:F_QS} are obtained using the expression for the curvature in an equilibrium field. The fourth term in \eqref{eq:df_QS1} is expressed in the following way upon application of \eqref{eq:delta_mod_B}, the divergence theorem, and noting that $\bm{\mathcal{S}} \cdot \nabla \psi = \nabla \cdot \bm{\mathcal{S}} = 0 $,
\begin{multline}
    \int_{V_P} d^3 x \, w(\psi) \mathcal{M} \bm{\mathcal{S}} \cdot \nabla \delta B = \int_{S_P} d^2 x \, \bm{\xi}_1 \cdot \mathbf{n}  B  w(\psi) \bm{\mathcal{S}} \cdot \nabla \mathcal{M}  - \int_{V_P} d^3 x \, \bm{\xi}_1 \cdot \left[B \nabla \left( w(\psi) \bm{\mathcal{S}} \cdot \nabla \mathcal{M} \right) \right]  \\
    + \int_{V_P} d^3 x \, w(\psi) (\bm{\mathcal{S}} \cdot \nabla \mathcal{M}) \left(\delta \iota_1(\psi) \textbf{b} \cdot(\nabla \psi \times \nabla \zeta)
    + B\bm{\xi}_1 \cdot \bm{\kappa} \right).
\end{multline}
We express terms involving $\bm{\xi}_1 \cdot \nabla$ as $\bm{\xi}_1 \cdot \nabla_{\perp}$ to obtain the corresponding terms in \eqref{eq:F_QS}. The fifth term in \eqref{eq:df_QS1} is expressed in the following way upon application of $\delta \psi = - \bm{\xi}_1 \cdot \nabla \psi$, the divergence theorem, and several vector identities,
\begin{multline}
    \int_{V_P} d^3 x \, w(\psi) \mathcal{M} \textbf{B} \times \nabla \delta \psi \cdot \nabla B = - \int_{S_P} d^2 x \, \bm{\xi}_1 \cdot \textbf{n}  w(\psi) \mathcal{M} \nabla B \times \textbf{B} \cdot \nabla \psi  \\
     - \int_{V_P} d^3 x \, \bm{\xi}_1 \cdot \nabla \psi \nabla B \cdot \nabla \times \left( w(\psi) \mathcal{M} \textbf{B} \right).
\end{multline}
The sixth term in \eqref{eq:df_QS1} upon application of \eqref{eq:delta_G} is,
\begin{multline}
-\int_{V_P} d^3 x \, \frac{\delta G(\psi)w(\psi) \mathcal{M} \textbf{B} \cdot \nabla B }{\iota(\psi)-(N/M)} = \\
     \frac{1}{4\pi^2} \int_{S_P} d^2 x \, \frac{w(\psi)  V'(\psi) \langle \mathcal{M} \textbf{B} \cdot \nabla B \rangle_{\psi}}{(\iota(\psi)-(N/M))} \left( \textbf{B} \cdot \nabla \psi \times \nabla \theta \right) \bm{\xi}_1 \cdot \textbf{n} \\
    - \frac{1}{4\pi^2} \int_{V_P} d^3 x \,  \bm{\xi}_1 \cdot  \nabla \left(\frac{w(\psi) V'(\psi) \langle \mathcal{M}\textbf{B} \cdot \nabla B \rangle_{\psi}}{(\iota(\psi)-(N/M))} \right) \textbf{B} \cdot \nabla \psi \times \nabla \theta \\
    + \frac{1}{4\pi^2} \int_{V_P} d^3 x \, \frac{w(\psi)V'(\psi) \langle \mathcal{M} \textbf{B} \cdot \nabla B \rangle_{\psi}}{\iota(\psi)-(N/M)} \left( \bm{\xi}_1 \cdot \left(\nabla \psi \nabla \cdot \left( \textbf{B} \times \nabla \theta \right) -\textbf{B} \times \nabla \times \left(\nabla \psi \times \nabla \theta \right)\right) \right)\\
    - \frac{1}{4\pi^2}\int_{V_P} d^3 x \, \delta \iota_1(\psi) \frac{w(\psi)V'(\psi) \langle \mathcal{M} \textbf{B} \cdot \nabla B \rangle_{\psi}}{\sqrt{g}^2(\iota(\psi)-(N/M))}  \partder{\textbf{r}}{\zeta} \cdot \partder{\textbf{r}}{\theta}. 
\end{multline}
In obtaining the corresponding terms in \eqref{eq:F_QS}, terms involving $\bm{\xi}_1 \cdot \nabla$ are expressed as $\bm{\xi}_1 \cdot \nabla_{\perp}$.
The seventh term in \eqref{eq:df_QS1} is expressed using $\delta \psi =- \bm{\xi}_1 \cdot \nabla \psi$. Combining all terms, we obtain \eqref{eq:df_QS}-\eqref{eq:I_QS}.

\section{Details of neoclassical figures of merit calculation}
\label{app:nc}

In this Section we compute the shape derivative of $f_{NC}$ \eqref{eq:f_NC} to obtain \eqref{eq:df_NC}-\eqref{eq:I_NC} by expressing each term in \eqref{eq:deltaf_NC} in the desired form. Throughout Boozer coordinates will be assumed.  

The second term in \eqref{eq:deltaf_NC} is expressed using $\delta \psi = - \bm{\xi}_1 \cdot \nabla \psi$. The third term in \eqref{eq:deltaf_NC} can be computed using \eqref{eq:delta_G}, noting that $V'(\psi)/(4\pi^2\sqrt{g}) = B^2/\langle B^2 \rangle$ in Boozer coordinates and applying the divergence theorem,
\begin{multline}
    \int_{V_P} d^3 x \, w(\psi) \partder{\mathcal{R}(\psi)}{G(\psi)} \delta G(\psi) = -\int_{V_P} d^3 x \, w(\psi) \frac{B^2 \sqrt{g}}{\langle B^2 \rangle_{\psi}} \partder{\mathcal{R}(\psi)}{G(\psi)} \bm{\xi}_1 \cdot \nabla \psi (\nabla \times \textbf{B}) \cdot \nabla \theta \\
    + \int_{V_P} d^3 x \,  \left( \bm{\xi}_1 \cdot \nabla \left( \partder{\mathcal{R}(\psi)}{G(\psi)} \frac{w(\psi)}{\langle B^2 \rangle_{\psi}} \right)B^2 G(\psi) + \frac{w(\psi)}{\langle B^2 \rangle_{\psi}}  \partder{\mathcal{R}(\psi)}{G(\psi)} \bm{\xi}_1 \cdot \textbf{B} \times \nabla \times \left( \partder{\textbf{r}}{\zeta}B^2\right) \right)  \\
    + \int_{V_P} d^3 x \,  \frac{w(\psi)\delta \iota_1(\psi)B^2}{\sqrt{g}\langle B^2 \rangle_{\psi}} \partder{\mathcal{R}(\psi)}{G(\psi)} \partder{\textbf{r}}{\zeta} \cdot \partder{\textbf{r}}{\theta} 
    - \int_{S_P} d^2 x \, w(\psi) \frac{B^2}{\langle B^2 \rangle_{\psi}} \partder{\mathcal{R}(\psi)}{G(\psi)} G(\psi)\bm{\xi}_1 \cdot \textbf{n}  .
\end{multline}
The fifth term in \eqref{eq:deltaf_NC} can be computed using \eqref{eq:delta_mod_B}, the divergence theorem, and the expression for the curvature in an equilibrium field,
\begin{multline}
    \int_{V_P} d^3 x \, w(\psi) \langle S_{\mathcal{R}} \delta B \rangle_{\psi} = \int_{V_P} d^3 x \, \left(\bm{\xi}_1 \cdot \nabla \left( w(\psi) S_{\mathcal{R}}\right) B - B S_{\mathcal{R}} w(\psi) \bm{\xi}_1 \cdot \bm{\kappa} \right) \\
  -\int_{V_P} d^3 x \, \delta \iota_1(\psi) S_{\mathcal{R}}w(\psi) \textbf{b} \cdot \nabla\psi \times \nabla \zeta - \int_{S_P} d^2 x \, w(\psi) S_{\mathcal{R}} B \bm{\xi}_1 \cdot \textbf{n}. 
\end{multline}
 The resulting terms can be combined to write the shape derivative in the form of \eqref{eq:df_NC}, noting that any terms involving $\bm{\xi}_1 \cdot \nabla$ can be expressed as $\bm{\xi}_1 \cdot \nabla_{\perp}$.

\end{document}